\documentclass[a4paper,fleqn,usenatbib]{mnras}

\usepackage[T1]{fontenc}
\usepackage{ae,aecompl}

\usepackage{graphicx}	% Including figure files
\usepackage{amsmath}	% Advanced maths commands
\usepackage{amssymb}	% Extra maths symbols

\usepackage{newtxtext,newtxmath}
\newcommand{\se}[1]{\S\ref{sec:#1}}

\newcommand{\Fig}[1]{Figure~\ref{fig:#1}}
\newcommand{\Figs}[1]{Figures~\ref{fig:#1}}
\newcommand{\tab}[1]{Table~\ref{tab:#1}}
\newcommand{\be}{\begin{equation}}
\newcommand{\ee}{\end{equation}}
\newcommand{\bea}{\begin{eqnarray}}
\newcommand{\eea}{\end{eqnarray}}

\newcommand{\msun}{{\rm M}_\odot}
\newcommand{\Msun}{M_\odot}

\newcommand{\ifm}[1]{\relax\ifmmode#1\else$\mathsurround=0pt #1$\fi}
\newcommand{\kms}{\ifmmode\,{\rm km}\,{\rm s}^{-1}\else km$\,$s$^{-1}$\fi}

\newcommand{\kpc}{\,{\rm kpc}}
\newcommand{\pc}{\,{\rm pc}}

\newcommand{\Myr}{\,{\rm Myr}}

\newcommand{\ltsima}{$\; \buildrel < \over \sim \;$}
\newcommand{\lsim}{\lower.5ex\hbox{\ltsima}}
\newcommand{\gtsima}{$\; \buildrel > \over \sim \;$}
\newcommand{\gsim}{\lower.5ex\hbox{\gtsima}}

\def\omm{\Omega_{\rm m}}
\def\oml{\Omega_{\Lambda}}
\def\omb{\Omega_{\rm b}}

\def\cmc{\,{\rm cm}^{-3}}
\def\cms{\,{\rm cm}^{-2}}

\def\Mv{M_{\rm vir}}
\def\Rv{R_{\rm v}}

\def\Ms{M_*}

\usepackage{color}

\title[VELA-6]{Effects of feedback on galaxies in the VELA simulations: elongation, clumps and compaction}

\author[Ceverino et al.]{
Daniel Ceverino$^{1,2}$\thanks{E-mail: daniel.ceverino@uam.es},
Nir Mandelker$^{3,4}$,
Gregory F. Snyder$^5$,
Sharon Lapiner$^3$, 
\newauthor
Avishai Dekel$^{3,6}$,
Joel Primack$^7$,
Omri Ginzburg$^3$, 
Sean Larkin$^7$
\\
$^{1}$Departamento de Fisica Teorica, Modulo 8, Facultad de Ciencias, Universidad Autonoma de Madrid, 28049 Madrid, Spain\\
$^{2}$CIAFF, Facultad de Ciencias, Universidad Autonoma de Madrid, 28049 Madrid, Spain\\
$^{3}$Centre for Astrophysics and Planetary Science, Racah Institute of Physics, The Hebrew University, Jerusalem 91904, Israel\\
$^{4}$Kavli Institute for Theoretical Physics, Kohn Hall, Santa Barbara, CA 93106, USA\\
$^{5}$Space Telescope Science Institute, 3700 San Martin Dr, Baltimore, MD 21218, USA\\
$^{6}$SCIPP, University of California, Santa Cruz, CA 95064, USA\\
$^{7}$Department of Physics, University of California, Santa Cruz, CA, 95064, USA \\
}

\date{Accepted XXX. Received YYY; in original form ZZZ}

\pubyear{2020}

\begin{document}
\label{firstpage}
\pagerange{\pageref{firstpage}--\pageref{lastpage}}
\maketitle

\begin{abstract}
The evolution of star-forming galaxies at high redshifts is very sensitive to the strength and nature of stellar feedback.
Using two sets of cosmological, zoom-in simulations from the VELA suite, we compare the effects of two different models of feedback:
with and without kinetic feedback
from the expansion of supernovae shells and stellar winds.
At a fixed halo mass and redshift, the stellar mass is reduced by a factor of $\sim$1-3 in the models with stronger feedback, so
the stellar-mass-halo-mass relation is in better agreement with abundance matching results.
On the other hand, 
the three-dimensional shape of low-mass galaxies is elongated along a major axis in both models.
At a fixed stellar mass, $\Ms<10^{10} \ \msun$, galaxies are more elongated in the strong-feedback case.
More massive, star-forming discs with high surface densities  form giant clumps.
However, the population of round, compact, old (age$_c > 300 \ {\rm Myr}$), quenched, stellar (or gas-poor) clumps  is absent in the model with strong feedback. On the other hand, giant star-forming clumps with intermediate ages (age$_c = 100-300 \ {\rm Myr}$) can survive for several disc dynamical times,
independently of feedback strength.
The evolution through compaction followed by quenching in the plane of central surface density and specific star-formation rate is similar under the two feedback models.
\end{abstract}


%%%%%%%%%%%%%%%%% BODY OF PAPER %%%%%%%%%%%%%%%%%%

%%%%%%%%%%%%%%%%%%%%%%%%%%%%%%%%%%%%%%%%%%%%%%%%%%
\section{Introduction}
%%%%%%%%%%%%%%%%%%%%%%%%%%%%%%%%%%%%%%%%%%%%%%%%%%

Galaxies at high redshifts, $z\ge1$, generate high levels of star formation, especially during the cosmic noon era, $z\simeq1-4$ \citep[e.g.,][]{Whitaker12, Madau14, FS20}.
The importance of stellar feedback from massive stars on galaxy-wide properties reaches a maximum during this epoch.
The injection of energy, momentum, mass and metals has important consequences for the self-regulation of star formation \citep{DekelSilk86, Governato07, Ceverino14, Hopkins14,Agertz15},  galactic outflows \citep{Oppenheimer08, Hopkins12a, Muratov15,Ceverino16}, and metal enrichment inside and outside galaxies \citep{Oppenheimer06, Ceverino16a, Langan20}. 
Other important phenomena related to galaxy formation at high-z may also be sensitive to feedback: galaxy elongation \citep[][C15 hereafter]{Ceverino15b}, the formation and evolution of giant clumps within massive  star-forming discs \citep[][M17 hereafter]{Mandelker17}, and the compaction and quenching of compact galaxies \citep[][Z15 hereafter]{Zolotov15}. These processes are described below.

Observations that estimate the intrinsic shapes of the stellar components of high-z galaxies find that the distribution of projected axis ratios of high-z samples at $z = 1.5 -4$ is inconsistent with a population of randomly oriented disc galaxies \citep{Law12, vdWel14, Zhang19}. They conclude that the intrinsic shapes of low-mass galaxies, $\Ms<10^{10} \ \Msun$, are strongly triaxial. Therefore, galaxies evolve from triaxial to oblate as they build their stellar mass. 
This build-up is regulated by feedback. Therefore, this morphological transformation may depend on the feedback strength. 
 
 From the theoretical side, elongated galaxies are predicted within the current $\Lambda$ cold dark matter ($\Lambda$CDM) paradigm \citep[C15,][]{Tomassetti16, Meng19, Pillepich19}. 
Triaxiality is a common property of dark matter (DM) halos in N-body-only simulations \citep[][ and references therein]{Jing02, Allgood06, Schneider12}.
A general result is that halos at a given mass scale are more prolate at earlier times. 
At the same time, the shape of the inner DM halo could influence the shape of the central galaxy \citep{DekelShlosman83} through the gravitational potential. If the galaxy is dominated by a non-axisymmetric, non-rotating potential, the stellar orbits are elongated along the major axis of the triaxial halo \citep[][sections 3.3.1 and 3.4.1]{BinneyTremaine}.
Therefore, the degree and evolution of elongation depend on the relative contribution of dark matter to the inner gravitational potential. This is related to the galactic stellar-to-DM fraction, which is sensitive to feedback. 

A large fraction of massive, baryon-dominated galaxies at high-z are confirmed to be rotating but turbulent discs \citep{Genzel06, FS09, Wisnioski15}. Many of these discs are broken into UV-bright giant clumps  that each account for a few percent of the disc mass and  at least 8\% of the UV light \citep{EE05, Genzel08, Guo15, Guo18}.
They are massive starburst regions, $\Ms\simeq 10^7 - 10^9 \ \Msun$, with stellar ages of around 100-200 Myr, consistent with recent episodes of star-formation \citep{Wuyts12}.
However observational estimates of the basic properties of clumps are still uncertain and they require a careful consideration of the systematic errors, using for example tools from machine learning \citep{Huertas20, Ginzburg21}.

According to our understanding of high-z galaxy formation, these turbulent discs suffer violent disc instabilities (VDI), in which the strong inflow of cold gas \citep{Dekel09} maintains high disc surface densities and marginally-unstable discs \citep{DSC}. In this scenario, the disc fragments into giant clumps around the Toomre mass scale \citep{Toomre, Inoue16}.
This has been simulated in idealized conditions \citep{Noguchi99, BournaudElmegreen09, Hopkins12b}, and in cosmological simulations of galaxy formation \citep{Agertz09, CDB}.

The lifetime of giant clumps is very sensitive to the strength and nature of stellar feedback. 
This has generated a debate about the final fate of these clumps.
Simulations of moderate feedback have shown a population of clumps that live for many orbital periods, with lifetimes upto 1 Gyr \citep[][M17]{CDB,Mandelker14}.
On the other hand,  simulations with stronger feedback disrupt clumps in much shorter times scales, 10-50 Myr \citep{Genel12,Oklopcic17,Buck17}.
These simulations use very different models of feedback, acting on different spatial and temporal scales. 
Therefore, their comparison is not straightforward. 

VDI and clump migration account for about 25\% of the total gas inflow to the galaxy center (Z15).
The rest of the flow is coming from wet mergers and counter-rotating streams.
If this inflow is much
higher than the disc star-formation rate (SFR), this dissipative process leads to wet compaction \citep{DekelBurkert}.
This is a characteristic evolution pattern of high-z galaxies, composed by different phases (Z15).
First, galaxies perturbed by gas-rich mergers, VDI or counter-rotating streams undergo a dissipative contraction into compact, star-forming galaxies \citep{Barro13} or massive bulges \citep{Luca21}.
The increase of the stellar surface density within the inner 1-kpc radius, $\Sigma_1$, occurs at a roughly time-independent specific SFR.
Secondly,  after compaction, there is a  inside-out quenching of star formation into a compact, quiescent galaxy \citep{Tacchella16b}.
This is driven by a fast gas consumption and strong outflows and it occurs at  constant $\Sigma_1$. Finally, a long-term suppression of gas supply is needed to maintain quenching, through halo quenching \citep{BirnboimDekel03} or AGN maintenance mode \citep{Croton06, Cattaneo09, Fabian12}.
These processes are all sensitive to the strength and timescales of stellar feedback.

%\textcolor{magenta}{\bf [NM: As it is, since it is not certain why these 2 specific feedback mechanisms are the 2 best to compare, and since the analysis of the different phenomena is not terribly detailed, I worry that the significance of this may be lost on some readers.]} 

%\textcolor{magenta}{\bf [AD: It is important to explain that feedback has many faces that we don't study them all. We use two simulation suites with two specific subgrid feedback models, and compare them, in order to get a feeling for the qualitative effects of the strenght of feedback.]} 

Stellar feedback is a mix of many different processes, from supernova explosions to radiation pressure.
By adding new mechanisms, we can increase the strength of feedback and
quantify its effect on the above processes of high-z galaxy formation:  galaxy elongation, giant clumps, and compaction. 
In this numerical experiment, we are going to compare two sets of cosmological simulations from the VELA suite with different models of feedback. 
This comparison can give us some insights about the trends with feedback strength.

Our first set of 35 simulated galaxies to be studied in detail was introduced in \cite{Ceverino14} and Z15 and it is here renamed as VELA-3. The second set, dubbed VELA-6, includes the injection of momentum from the expansion of supernovae shells and stellar winds \citep{FirstLight}.
Section \se{runs} summarises both models in detail. 
Section \se{SMHM} describes the effect of feedback on the stellar-mass-halo-mass relation.
Sections \se{e}, \se{Clumps} and \se{C} are devoted to galaxy elongation, clumps and compaction respectively.
Finally, Section \se{summary} ends with the summary and discussion.

%%%%%%%%%%%%%%%%%%%%%%%%%%%%%%%%%%%%%%%%%%%%%%%%%%
\section{Simulations}
\label{sec:runs}
%%%%%%%%%%%%%%%%%%%%%%%%%%%%%%%%%%%%%%%%%%%%%%%%%%

\begin{table*} 
\caption{Properties of the main progenitors in VELA-3 and VELA-6 simulations at $z=2$ and $z_{\rm last}$, the last available redshift. The unit for virial radius, $\Rv$ is proper kpc. Virial masses, $\Mv$, have units of $10^{12} \ \msun$ and stellar masses, $\Ms$, have units of $10^{10} \ \msun$.  
The stellar mass is defined using all stars within a sphere of $0.1\Rv$.
At a fixed halo mass and redshift, the stellar mass is reduced by a factor of $\sim$1-3 in models with stronger feedback. } 

 \begin{center} 
 \begin{tabular}{ccccccccccccccc} \hline 
             & & & &   &  VELA-3 & & & & &  & &VELA-6 \\
Run  & $z_{\rm last}$ & $\Rv$ & $\Mv$ & $\Ms$ & $\Rv(z=2)$ & $\Mv(z=2)$ & $\Ms(z=2)$ & $z_{\rm last}$ & $\Rv$ & $\Mv$ & $\Ms$ & $\Rv(z=2)$ & $\Mv(z=2)$ & $\Ms(z=2)$ \\ 
V01 & 1 & 124 & 0.48 & 1.5 & 58 &  0.16 & 0.2 & 1 & 122 & 0.46 & 0.79 & 57 & 0.15 & 0.057 \\
V02 & 1 & 115 & 0.39 & 0.95&55 &  0.13 & 0.16&1 & 113 & 0.37 & 0.70 & 54 & 0.13 & 0.067 \\
V03 & 1 & 108 & 0.32 & 1.0 & 56 &  0.13 & 0.38& 1& 108 & 0.32 & 0.49 & 55 & 0.13 & 0.10 \\
V04 & 1 &  90 & 0.18 & 0.35& 53 &  0.12 & 0.082&1& 88 & 0.17 & 0.14 & 53 & 0.12 & 0.046 \\
V05 & 1 &  83 & 0.14 & 0.27& 44 & 0.070& 0.072&1& 83 & 0.15 & 0.095&43 & 0.063&0.032 \\
V06 &1.7&109 & 0.74 & 2.6  &88 & 0.54  & 2.1    &1&200& 1.9 & 4.4   & 93 & 0.63  &0.81 \\
V07 &1.5&145 & 1.4   & 7.3  &104 & 0.89  & 5.7   &1.5&150&1.5&  7.4   & 107 & 0.97  & 4.6 \\
V08 &0.9&182 & 1.3  &  4.4  &71 & 0.28  & 0.35 &0.9&183&1.3  &  2.9   &  71 & 0.28  & 0.11 \\
V09 &1.5&121 & 0.83&  4.3  &71 & 0.28  & 1.0   &0.8&209&1.7  &  5.1   & 64 & 0.21  & 0.50 \\
V10 &0.8&159 & 0.76&  3.4  &55 & 0.13  & 0.60 &0.8&182&1.14&  1.7   & 55 & 0.13  & 0.37 \\
V11 &0.8&106 & 0.37&  1.6  &70 & 0.27  & 0.76 &0.8&158&0.74&  0.73 & 68 & 0.25  & 0.37 \\
V12 &1.5& 86 & 0.29&  2.1  &69 & 0.26 &  1.9   &0.8&159&0.76& 2.8    & 76 & 0.35  & 1.5 \\
V13 &1.5&108 & 0.59&  2.2  &73 & 0.30 &  0.57 &0.8&169&0.92& 3.5    & 72 & 0.29  & 0.30 \\
V14 &1.8& 86 & 0.39&  2.3  &76 & 0.36 &  1.3   &0.8&145&0.58& 2.3    & 76 & 0.34  & 0.47 \\
V15 &0.8&123 & 0.35&  1.5  &53 & 0.12 &  0.51 &0.8&125&0.37& 0.52  & 53 & 0.12  & 0.18 \\ 
V16 &3.2& 63 & 0.50&  4.1    &-   & -     &  -        &2.3&  84&0.62& 3.6    & - & -  & - \\
V17& 2.2& 106 & 1.1&  8.4   & -   & -     & -         &2.2& 108&1.2 & 4.7    & - & - & -  \\
V18&-     &-      & -    &  -       &-     & -    &  -        &2.2&  79 & 4.8& 2.2    & - & -  & - \\
V19& 2.4& 91 & 0.87 &  4.4  & -    & -   &  -         &2.3&  96 & 0.92& 4.1  & - &  -    & - \\
V20 &1.3&146 & 1.1&  7.5  &87 & 0.53 &  3.5 &1.3&148&1.1& 5.6  & 101 & 0.82  & 4.0 \\ 
V21 &  1 &151 &0.88& 6.5  &92 & 0.62 &  4.0 &  1 &156&0.98&5.7 & 94 & 0.65  & 3.3 \\
V22 &  1 &136 &0.63& 4.6  &85 & 0.50 &  4.4 &  1 &150&0.86&4.4 & 88 & 0.54  & 3.4 \\
V23 &  1 &123 &0.47& 2.5  &57 & 0.15 & 0.76&  1 &122&0.46&1.4 & 57 & 0.15  & 0.47 \\
V24 & 1.1&108&0.36& 2.1  &70 & 0.28 & 0.87&  1 &117&0.41&1.5 & 69 & 0.27  & 0.65 \\
V25 &  1  &108&0.32& 1.4  &65 & 0.22 & 0.68&  1 &107&0.31&0.62&68 & 0.25  & 0.19 \\ 
V26 &  1  &120&0.44& 2.2  &77 & 0.36 & 1.6  &  1 &120&0.44&1.6  &76 & 0.35  & 1.1 \\
V27 &  1  &114&0.38& 2.0  &75 & 0.34 & 0.71&  1 &114&0.37&1.3  &75 & 0.34 &  0.48 \\
V28 &  1  & 96&0.22&0.49 &63 & 0.20 & 0.18&  1 &94 &0.21&0.30&63 & 0.20 & 0.082 \\
V29 &  1  &152&0.90&0.37 &84 & 0.52 & 0.20&  1 &153&0.91&3.2  &87 & 0.52 & 1.8 \\
V30 & 1.9& 76 & 0.32& 1.6 &73 & 0.31 & 1.6    &1.8&80&0.34&1.2  & 73 &  0.32    & 1.1 \\
V31& 4.3 & 38 & 0.22& 0.8 &-    & -       & -        &1.8&117&1.1 &4.1 & 105 &  0.93    & 3.8 \\
V32& 2   &  90 & 0.6  & 2.7 &90 & 0.6    & 2.7    &2  &101& 0.82& 2.2& 101 &  0.82    & 2.2 \\
V33& 1.6 &144& 1.5 &  9.4 &101 & 0.83   & 4.8  &1.5&148&1.5 &  6.9& 107 &  0.98    & 2.5 \\
V34& 1.8 &97  & 0.62& 1.9 &  86 & 0.34   & 0.76  &2.4 &66& 0.34& 0.32  &-   & -          & - \\
V35& 3.5 &37 & 0.14&  0.49 &-     & -        &  -      &3    &48&  0.2 & 0.23   & - & -          & - \\
\hline 
 \end{tabular} 
 \end{center} 
\label{tab:vela} 
 \end{table*} 

\begin{figure*}
	\includegraphics[width=2 \columnwidth]{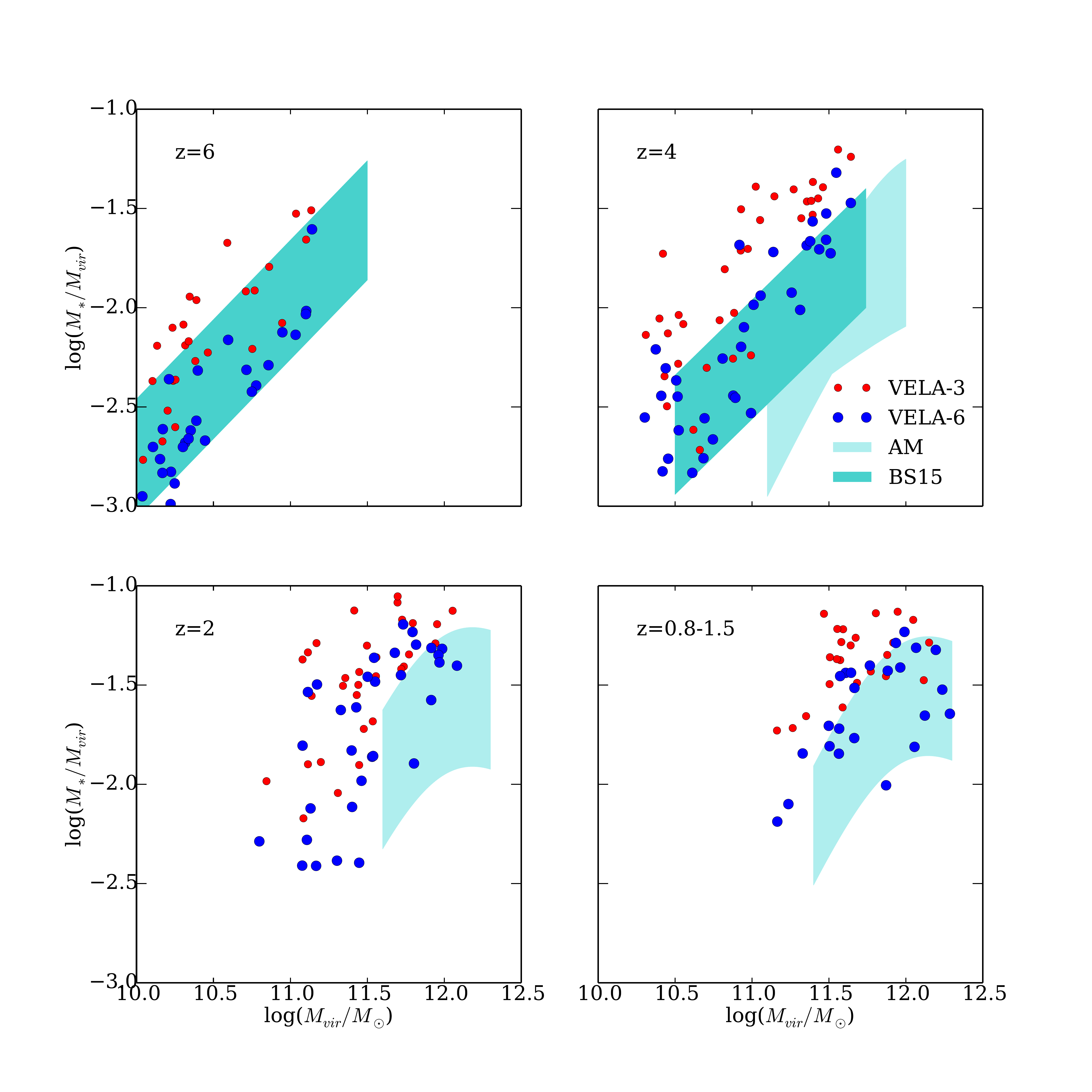}
		 \caption{Ratio between stellar mass and virial mass at different redshifts. At the same virial mass,  VELA-6 (blue) have a factor $\sim$1-3 lower stellar mass than VELA-3  (red), in better agreement with the results from abundance matching \citep[AM:][]{Moster13, Behroozi13, RodriguezPuebla17, Moster18, Behroozi19} and other empirical models at high redshift \citep[BS15:][]{BehrooziSilk15}.}
	  \label{fig:SMHM}
\end{figure*}

\begin{figure}
	\includegraphics[ trim={60 18 100 80},clip, width=\columnwidth]{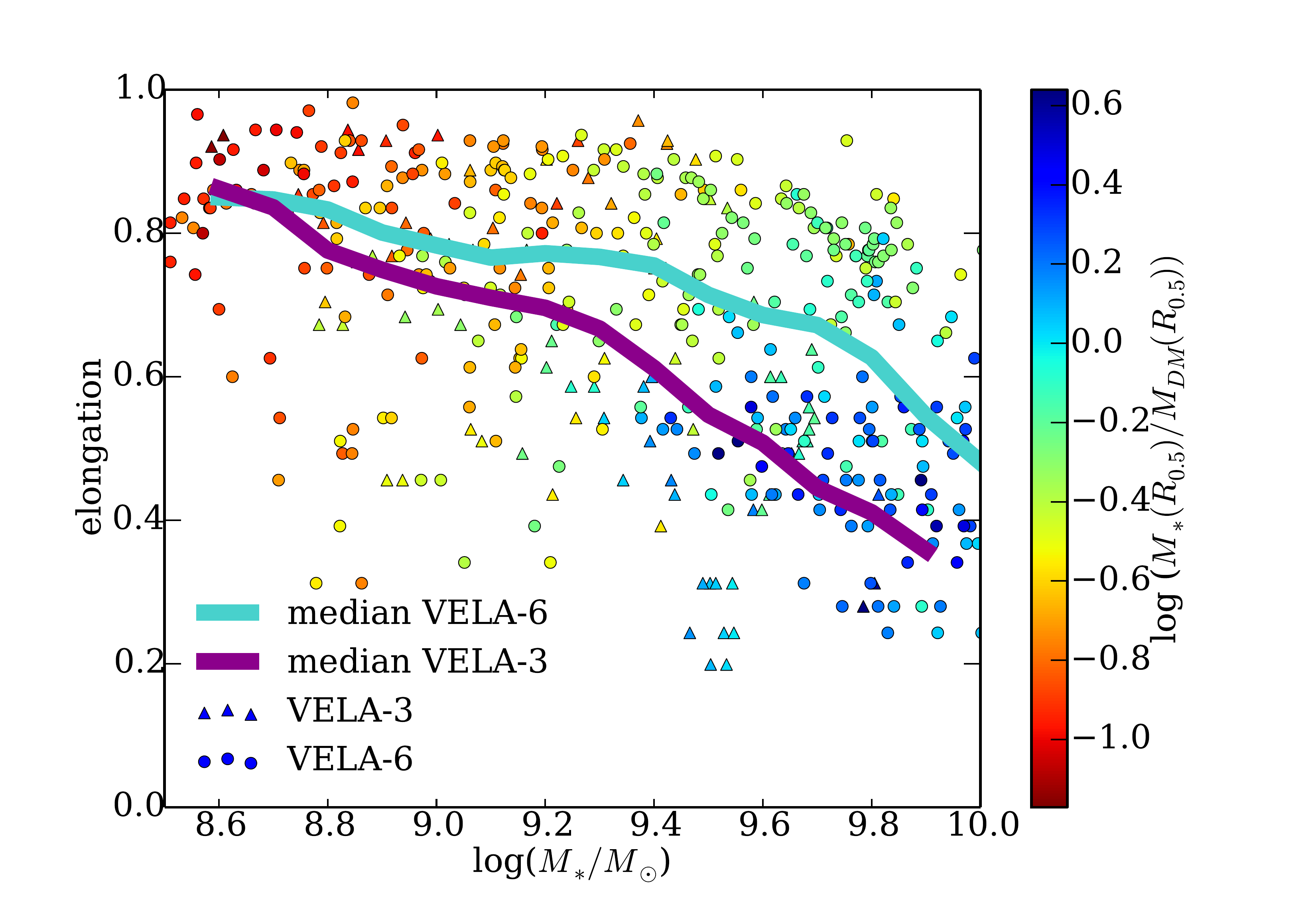}
		 \caption{Elongation parameter (eq. 1) versus stellar mass for all available snapshots of VELA-3 and VELA-6 at $z=1-3$, color coded by the star-to-DM mass ratio within the half-mass radius. The median values (solid lines) show that the elongation decreases with increasing mass and mass ratio but the trend is weaker for VELA-6. 
		 Galaxies  in models with stronger feedback are more elongated.
		 More massive galaxies ($\Ms>10^{10} \Msun$) are not shown as they are preferentially axialsymmetric.
		 }
	  \label{fig:e}
\end{figure}		

\begin{figure}
	\includegraphics[ trim={10 15 12 15},clip, width= \columnwidth]{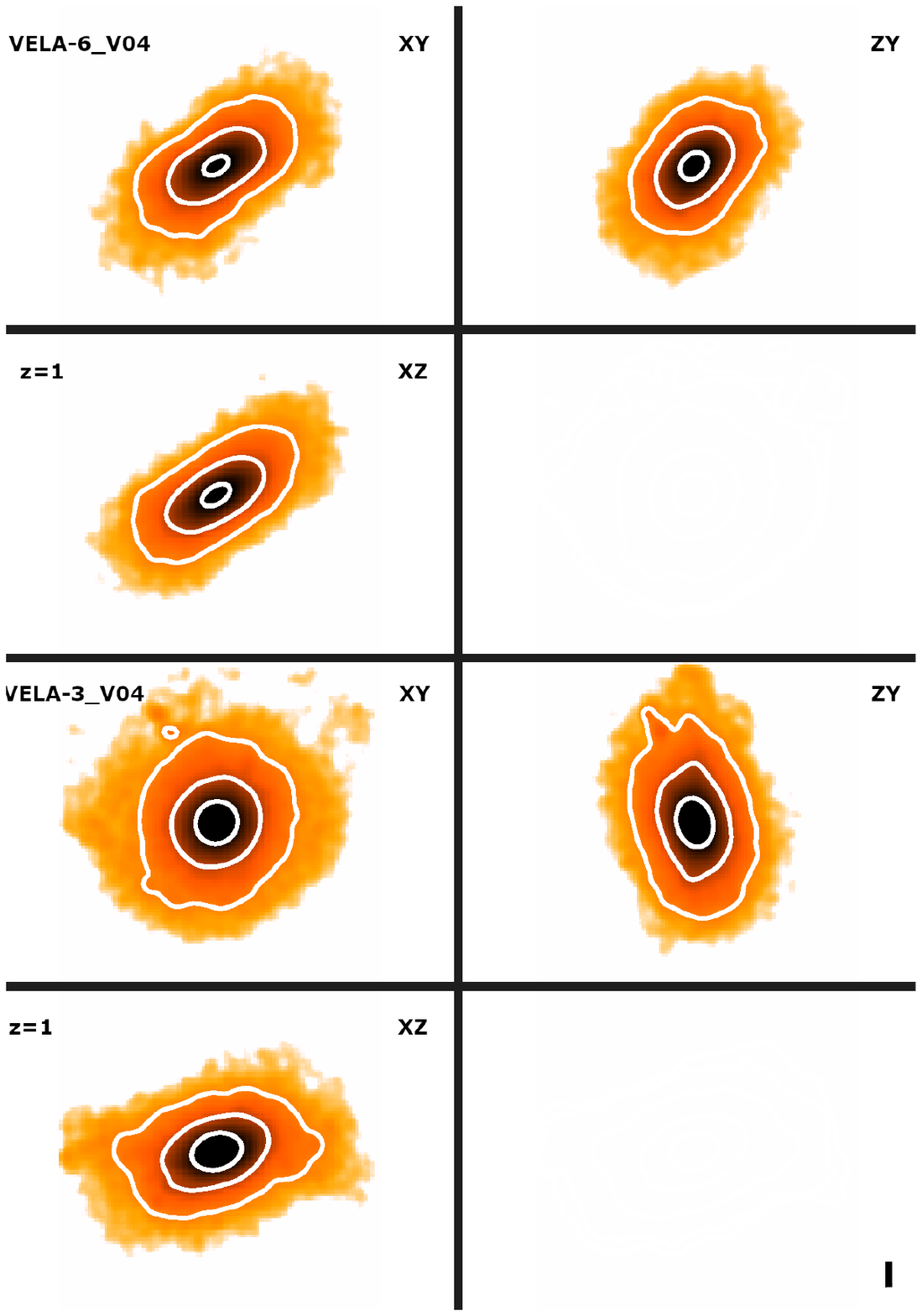}
		 \caption{ Three orthogonal projections (XY, XZ and ZY) of the stellar surface density of the same galaxy (V04) in the VELA-6 (top two panels) and VELA-3 (bottom two panels) at $z=1$. Contours show isodensities of 4, 20 and 100 $\Msun \pc^{-2}$. The bar in the bottom-right corner marks 1 kpc. VELA-6 galaxies are significantly more elongated, $a>b$, than the VELA-3 counterparts.}
	  \label{fig:examplesE}
\end{figure}	

\begin{figure*}
 	\includegraphics[width=2 \columnwidth]{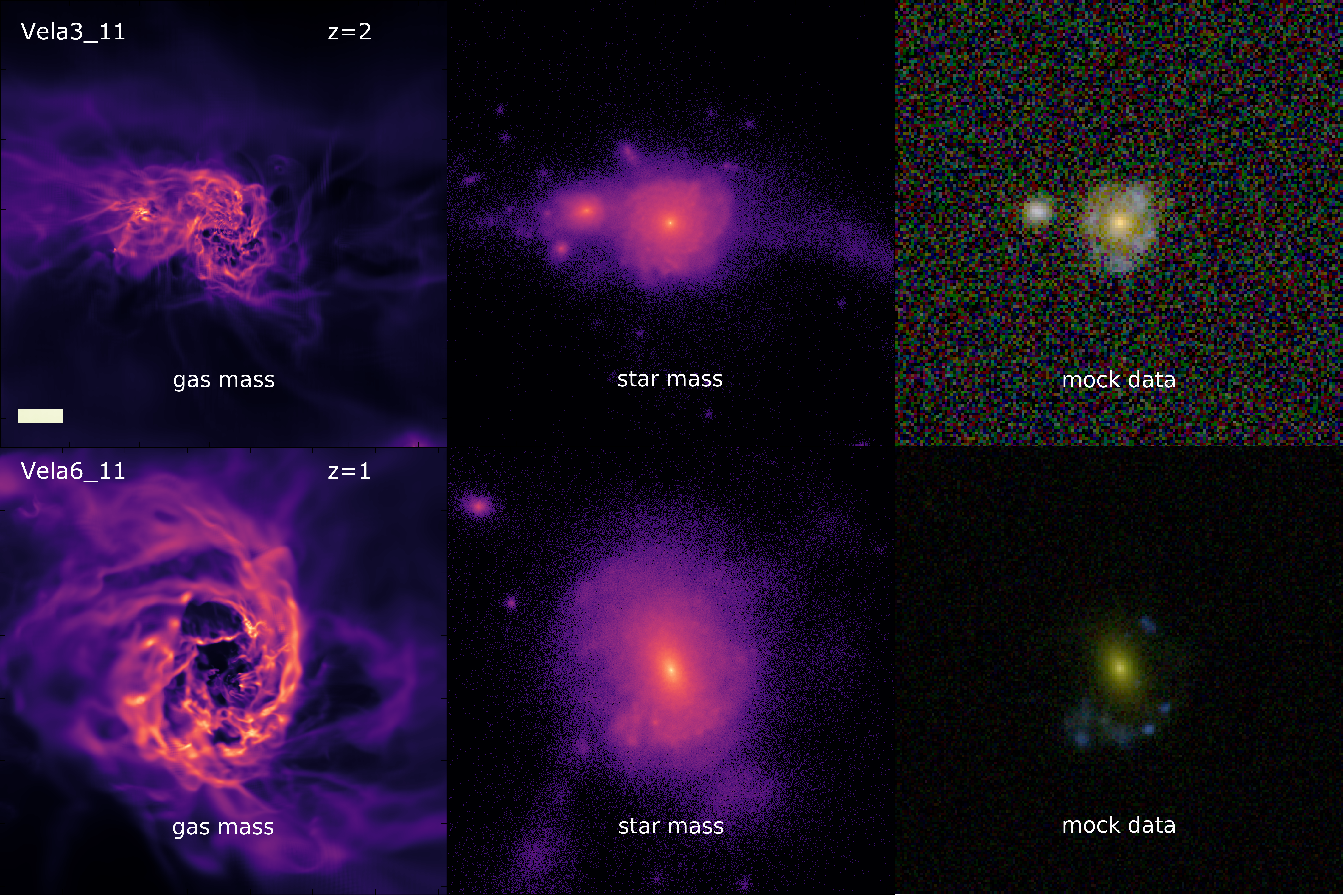}
		\caption{Two examples of face-on views of the same galaxy  (V11) in VELA-3 (top) and VELA-6 (bottom) datasets with the same stellar mass,  $\Ms\simeq 6 \times 10^9 \ \msun$, but at two different redshifts.
		The VELA-3 example shows a small axysimmetric disc while the VELA-6 case shows an elongated galaxy surrounded by a star-forming ring.
		The size of the images is 100 kpc. The scale bar marks 10 kpc.
		The mock HST images (right) are roughly rest-frame U-B-V bands.} 
	  \label{fig:sunrise}
\end{figure*}

The VELA-3 and VELA-6 simulations use the same set of initial conditions. 
 They only differ in their models of stellar feedback.
 These models have been developed from a pure thermal feedback model (VELA-2) with the addition of radiation pressure (VELA-3), photoheating/photoionization (VELA-4), and a moderate trapping of infrared photons (VELA-5). The last model, VELA-6, includes thermal, radiation pressure, moderate IR trapping and kinetic feedback. 
 The comparison between VELA-2 and VELA-3 was discussed in \cite{Ceverino14}, \cite{Moody14} and M17.
 
VELA-3 has been our standard set of simulations for several years.
In addition to the works already mentioned, these simulations have been used to study metallicity inhomogeneities within galaxies  \citep{Ceverino16a},  the velocities of warm outflows \citep{Ceverino16}, and the link between halo and galaxy spin \citep{Jiang19}, among other works that complement this paper. 
For example, VELA simulations
help to distinguish  between mergers and discs by simulating kinematic maps \citep{Simons19} and they also help to check the robustness of different dynamical mass estimates \citep{Kretschmer21}.
We learn how feedback regulates galaxy-wide star formation \citep{Dekel19},  the mass threshold above which galactic discs may form \citep{Dekel20a}, and the origin of star-forming rings \citep{Dekel20b}. 
Finally, the connection between  galaxy properties and the circungalactic medium has been described in \cite{Roca19, Strawn21}.
 The comparison between VELA-3, VELA-4 and VELA-5 yields very similar results.
 Therefore, in this paper we are going to focus on the comparison between VELA-3 and VELA-6.

\subsection{The Initial Conditions of VELA}

VELA halos were drawn from N-body simulations with cosmological boxes of 10, 20 or 40 Mpc/h across, assuming a
$\Lambda$CDM cosmology with $\omm=0.27$, $\oml=0.73$, $\omb= 0.045$, $h=0.7$ and $\sigma_8=0.82$.
They were selected randomly in log $\Mv$ in order to cover a halo mass range between $\Mv \sim10^{11}$ and $\sim10^{12} \ \msun$ at $z=1-0.8$. Only ongoing major mergers at $z=1$ were excluded. This only discarded 10\% of the halos.
In isolation, the median mass at $z = 0$ is about  $10^{12}\ \msun$. However, the mass range is broad, with some of the halos merging into more massive halos hosting galaxy groups at $z = 0$ (Z15).

The selected halos were filled with gas and refined to a much higher resolution on an adaptive
mesh within a zoom-in lagrangian volume that encompasses the mass within
twice the virial radius \citep{Klypin02}.
The initial conditions of these runs contain between 6.4 to 46 $ \times 10^6$ dark matter particles 
 with a minimum mass of 
$8.3 \times 10^4 \ \msun$, while the particles representing single stellar populations that were formed in the simulation
have a minimum mass of $10^3 \ \msun$. 
Each AMR cell is split into eight cells once it contains a mass in stars and dark matter higher than $2.6 \times 10^5 \ \msun$, equivalent to three dark matter particles, or once it contains a gas mass higher than $1.5 \times 10^6 \ \msun$. This quasi-Lagrangian strategy ends at the highest level of refinement that marks the minimum cell size at each redshift. This size is between 17-35 proper pc.
More details can be found in \citet{Ceverino14} and \citet{Zolotov15}.

\subsection{ART}
 
The  \textsc{ART} code \citep{Kravtsov97,Kravtsov03} accurately follows the evolution of a
gravitating N-body system and the Eulerian gas dynamics using an AMR approach.
Besides gravity and hydrodynamics, the code incorporates 
many of the astrophysical processes relevant for galaxy formation.  
These processes, representing subgrid 
physics, include 
gas cooling due to atomic hydrogen, helium,  molecular 
hydrogen and metals down to a minimum temperature of 300 K.
The code also includes a pressure floor to prevent artificial fragmentation \citep{CDB}.
Other relevant processes are photoionisation heating by a constant cosmological UV background with partial 
self-shielding, star formation and feedback, as described in 
\citet{Ceverino09}, \citet{CDB}, and \citet{Ceverino14}. 

\subsection{Thermal Feedback}

The model of thermal feedback assumes that each stellar particle acts as a single stellar population and injects their mechanical luminosity from supernovae and stellar winds as thermal heating, injected into the cell that host the stellar particle.
The model assumes a constant heating rate over 40 Myr, following the values from STARBURST99 \citep{Leitherer99}. 
More details and discussions can be found in \citet{Ceverino09}.

\subsection{Radiative Feedback}

In addition to thermal-energy feedback, the simulations use radiative feedback \citep{Ceverino14}.
This model adds a radiation pressure to the total gas pressure in regions where ionising photons from massive stars are produced and trapped. 
In the model used in VELA-3, named RadPre in \citet{Ceverino14}, radiation pressure is included in the cells (and their closest neighbours within a sphere of radius equal to the cell size)
that contain stellar particles younger than 5 Myr and whose gas column density exceeds $10^{21}\ \cms.$ 
Instead, VELA-6 uses the model RadPre\_IR.
In addition to the effect of ionising photons, the model adds
a  moderate trapping of infrared photons, only if the gas density in the host cell exceeds a threshold of 300 $\cmc$.
In this simple model, we assume that the optical depth of infrared photons scales linearly with the gas density, regardless of gas metallicity or dust composition.
More details and discussion can be found in \citet{Ceverino14}.

\subsection {Kinetic feedback}
\label{sec:ST}

In addition to radiative feedback, VELA-6 also includes the injection of momentum coming from the (unresolved) expansion of gaseous shells from supernovae and stellar winds \citep{OstrikerShetty11}, as described in \citet{Ceverino17b}.
A momentum of $10^6 \ \msun \kms$ per massive star (i.e per star more massive than 8 $\msun$) is injected at a constant rate over 40 Myr, the lifetime of the lightest star that explodes as a core-collapsed supernovae.
The model also takes into account a factor 3 boost in the injected momentum due to the clustering of supernovae \citep{Gentry17}.
The injection of momentum is implemented in the form of a non-thermal pressure, as described in \citet{Ceverino17b}.

%This model is similar to the injection of momentum described in \citet{Agertz13}.
This model differs from other recent implementations of kinetic feedback. It goes beyond the thermal-only feedback \citep{Stinson13, Schaye15}, and it does not shut down cooling in the star-forming regions \citep{Stinson06}. It does not impose a wind solution \citep{Hopkins14, Vogelsberger14}, so that outflows are generated in a self-consistent way \citep{Ceverino16}. Our implementation is more similar to the feedback model in  \citep{Agertz15}.

%%%%%%%%%%%%%%%%%%%%%%%%%%%%%%%%%%%%%%%%%%%%%%%%%%
\section{Stellar-Mass-Halo-Mass Relation}
\label{sec:SMHM}
%%%%%%%%%%%%%%%%%%%%%%%%%%%%%%%%%%%%%%%%%%%%%%%%%%

 One of the main effects of feedback is the self-regulation of the galaxy formation efficiency \citep[][and references therein]{Ceverino14, Hopkins14}.
That efficiency can be quantified by the stellar-mass-halo-mass (SMHM) relation.
The halo mass is measured as the virial mass \citep{BryanNorman} and the stellar mass is computed within 10\% of the virial radius.
\tab{vela} shows these properties at $z=2$ and at the last available snapshot for VELA-3 and VELA-6 respectively.
\Fig{SMHM} shows the SMHM ratio at different redshifts.
In general, that ratio decreases towards low masses, as stellar feedback regulates star formation.
There is a large scatter of about a factor 2 around the average ratio, driven by the different mass accretion histories \citep{Moster18}.
At a fixed halo mass,
VELA-6 halos have a factor $\sim$1-3 lower stellar mass  than in VELA-3, due to their stronger feedback.
The biggest difference is concentrated at low masses, $\Mv\le 10^{11} \ \msun$, where stellar feedback is most efficient \citep{DekelSilk86}.

VELA-6 galaxies agree better with available results from models of abundance matching \citep{Moster13, Behroozi13, RodriguezPuebla17, Moster18, Behroozi19}, and other empirical models at high redshifts \citep{BehrooziSilk15, Tacchella18}.
We added an uncertainty of a factor 2 in these models that takes into account systematic differences due to different cosmology and IMF assumptions, as well as uncertainties in the stellar mass estimates \citep{Leja19}.

VELA-6 also agrees with the FirstLight simulations at $z\simeq6$ \citep{Ceverino17b}, which use a similar feedback model but have twice better spatial resolution. These overall results seem numerically converged to the extent to which we are able to test.
They are still sensitive to feedback.
Only future simulations will tell us whether we have reached model-independent values.

%%%%%%%%%%%%%%%%%%%%%%%%%%%%%%%%%%%%%%%%%%%%%%%%%%
\section{Elongation}
\label{sec:e}
%%%%%%%%%%%%%%%%%%%%%%%%%%%%%%%%%%%%%%%%%%%%%%%%%%

\begin{figure*}
	\includegraphics[width=2 \columnwidth]{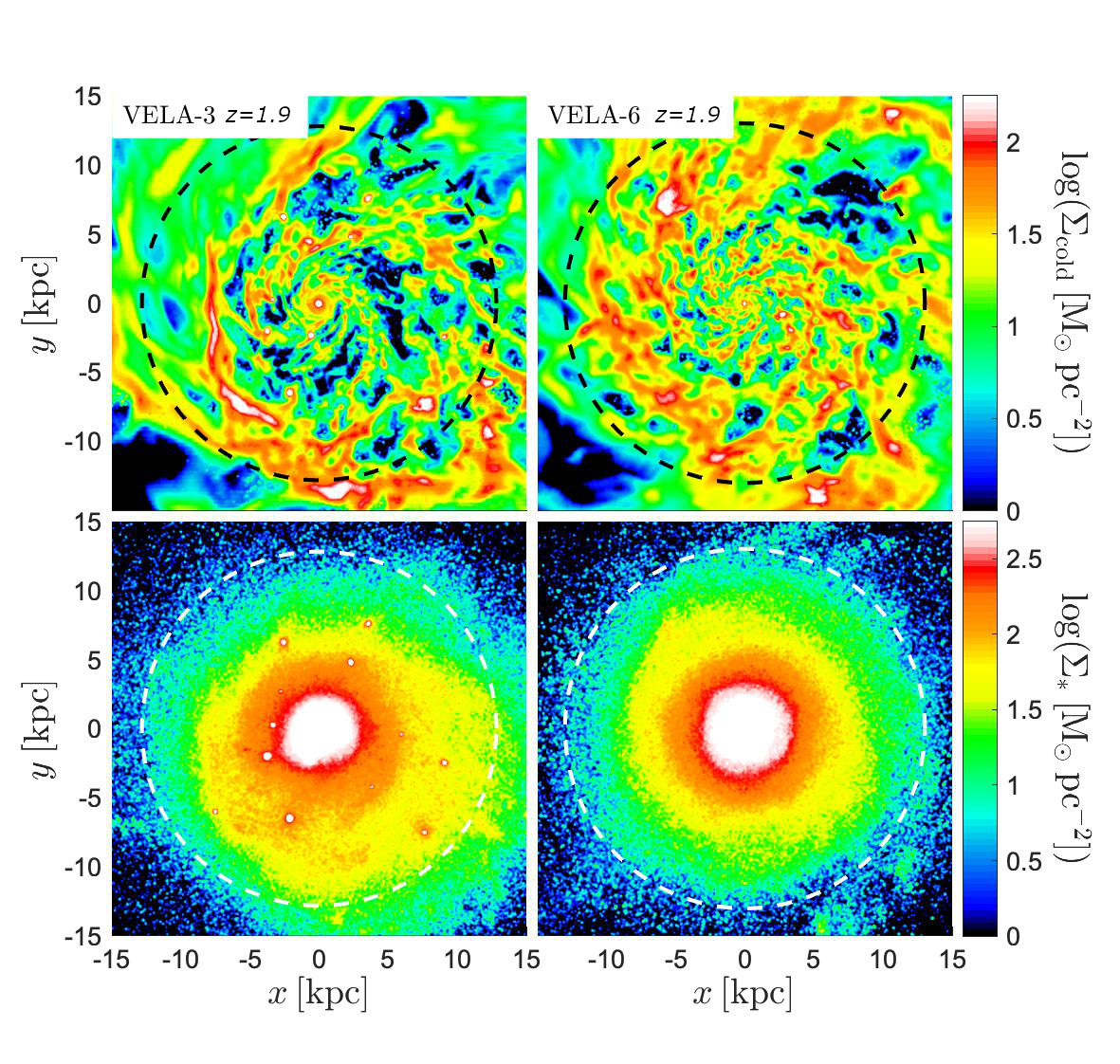}	
		\caption{Face-on views of cold gas plus young stars, T$<1.5 \times 10^4$ K and $ \rm{age}<100 \Myr$ (top) and all stars (bottom) of V07 at $z\simeq2$ for VELA-3 (left) and VELA-6 (right). Circles represent the disc radius as defined in M17.
		Clumps in VELA-3 are significantly denser and rounder.
		Gas is clumpy in both models but dense clumps are absent in the stellar map of VELA-6.} 
	  \label{fig:clumps}
\end{figure*}	
\begin{figure}
	\includegraphics[width=\columnwidth]{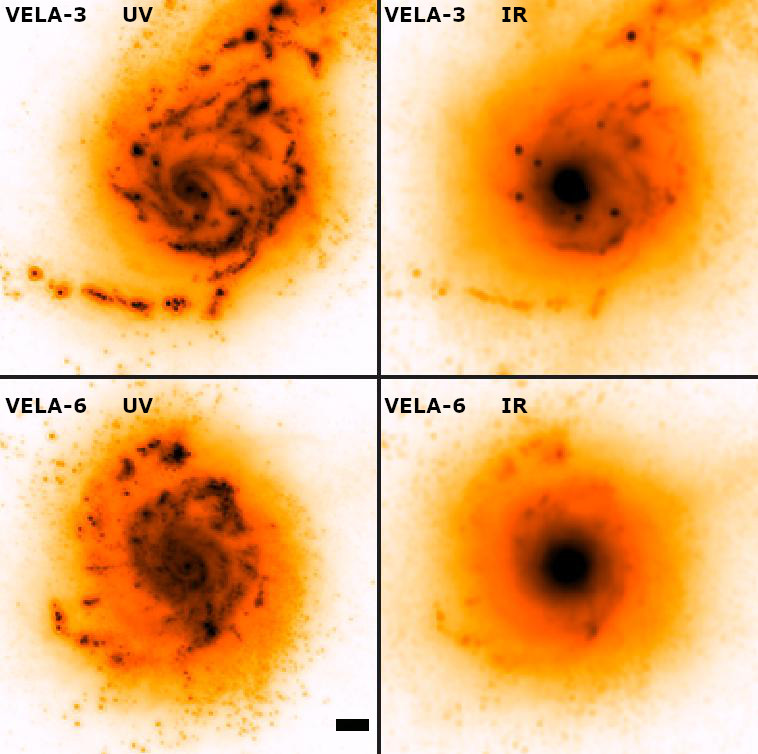}
		\caption{Mock images in the HST F606W band (left) and JWST F277W band (right) for VELA-3 (top) and VELA-6 (bottom) of the same galaxy from \Fig{clumps}, assuming a Milky-Way-like dust model. The bar marks 5 kpc. The rest-frame UV is clumpy in both models. However, the rest-frame near-IR is significantly  less clumpy  in models with stronger feedback.}
	  \label{fig:obs}
\end{figure}

The degree of intrinsic, 3D elongation of a galaxy can be quantified by the elongation parameter, as defined in C15,
\begin{equation} 
e= \sqrt{ 1 - (b/a)^2 }
\end{equation}
where $a$ and $b$ are the  major and intermediate axes of a 3D ellipsoidal shell that fits the stellar iso-density surface at the galaxy 3D half-mass radius, $R_{0.5} \simeq a$, as described in C15.
In short, the calculation starts by computing the stellar density at the point of each stellar particle by using a 80-particle kernel. 
A first guess of the major axis direction at a radius $R_{0.5}$ uses the eigenvectors of the inertia tensor computed from all stars within a spherical shell at that radius. Then, the isodensity surface is defined by all particles with assigned density similar to the density at $R_{0.5}$ in the direction of the major axis.
Finally, an ellipsoidal fit to that isodensity surface defines the main axes: $a$, $b$ and $c$.

\Fig{e} shows elongation versus galaxy mass, coloured by the stellar-to-DM mass ratio within $R_{0.5}$.
The results of VELA-3 were obtained in C15.
These simulations show that
a galaxy achieves a high elongation if DM mass dominates over stellar mass within the galaxy.
In this case, the elongation of the DM halo dominates the gravitational potential, which sets the galaxy elongation. 
As galaxy mass and halo mass increases, the stellar fraction increases and the galaxy elongation decreases. 
This is because baryons make rounder halos \citep{Zemp12}, especially in the inner parts, where baryons dominate the potential.
VELA-6 shows a weaker trend between elongation and mass, because star formation is less efficient at a fixed halo mass.

At a fixed stellar mass, the mass fraction in DM is higher in VELA-6. Therefore, they are more DM dominated and more elongated than VELA-3.
That difference is more striking for masses between $10^{9.5}$ and $10^{10} \ \msun$, where galaxies are preferentially more elongated in VELA-6 than in VELA-3, independently of the actual value of the elongation parameter.

\Fig{examplesE} shows stellar mass projections of the same galaxy (V04) at $z=1$ in both suites. The XY projection coincides with the face-on view of the galaxy, according to the angular momentum of the cold gas.
The VELA-6 case shows a typical elongated galaxy ($e=0.7$) of low stellar mass, $\Ms=1.4 \times 10^9 \ \Msun$.
It is DM dominated within $R_{0.5}$,  $\Ms(R_{0.5})/M_{\rm DM}(R_{0.5})=0.25$.
This is similar to the case shown in C15.
The VELA-3 example is more massive, $\Ms=3.5 \times 10^{9} \ \Msun$, and it shows a more oblate shape ($a\simeq b$), typical of a massive, baryon-dominated galaxy with  $\Ms(R_{0.5})/M_{\rm DM}(R_{0.5})=1.2$.

\Fig{sunrise} compares the same galaxy in the two datasets (V11) but now the galaxy mass is the same, $\Ms\simeq 6.5 \times 10^9 \ \msun$.
Therefore, the selected snapshots have different redshifts.
Both galaxies have roughly the same luminosity but the VELA-6 case lives in a halo that is twice more massive, $\Mv=5.6 \times 10^{11} \ \msun$, than in the VELA-3 example.
This drives a big difference in their stellar-to-DM mass fraction within $R_{0.5}$. 
The VELA-6 case is DM-dominated, $\Ms(R_{0.5})/M_{\rm DM}(R_{0.5})=0.5$, while the contribution of baryons is much more relevant in  the VELA-3 example  ($\Ms(R_{0.5})/M_{\rm DM}(R_{0.5})=1$).

The galaxy in VELA-3 shows an oblate axysimmetric disc, as also shown in \cite{Tomassetti16}.
This is seen both in the stellar maps as well as in the mock HST image, obtained with SUNRISE as described in \cite{Simons19} and \cite{Snyder15}.
The VELA-6 case is far from axysimmetry ($e=0.8$ at $R_{0.5}\simeq 3 \kpc $).
The mock data in rest-frame U-B-V bands shows an elongated galaxy surrounded by a clumpy star-forming ring. 
The rest-frame visible bands follow the elongated stellar shape while the rest-frame UV traces the star-forming gas clumps of the gaseous disc.

%%%%%%%%%%%%%%%%%%%%%%%%%%%%%%%%%%%%%%%%%%%%%%%%%%
\section{Clumps} 
\label{sec:Clumps}
%%%%%%%%%%%%%%%%%%%%%%%%%%%%%%%%%%%%%%%%%%%%%%%%%%

Massive,  $\Ms \geq 10^{10} \Msun$, galaxies develop a thick rotating disk, which undergoes violent disk instability \citep{DSC, CDB}  and the formation of giant in-situ clumps \citep[][M17]{Mandelker14}.
We have used the clump finder described in M17 to generate catalogs of clumps in VELA-3 and VELA-6.
The finder uses a cloud-in-cell interpolation to deposit mass from star particles and cold ($T<1.5 \times 10^4$ K) gas cells  into a uniform 3D grid with a cell size of $\Delta=70$ pc. For each component, it applies  a gaussian filter to smooth the density field and it calculates a density residual in each point, $\delta_{\rho} = ( \rho - \rho_{\rm w})/ \rho_{ \rm w}$, where $\rho$ and $\rho_{\rm w}$ are the original and smoothed fields. 
Finally, it takes the maximum of the two residual values (stars or gas) at each point.
We define clumps as connected regions containing at least 8 grid cells above a residual threshold, $\delta^{\rm min}_{\rho} =10$.
More details and a discussion of the sensitivity of results to parameter variations can be found in M17.
The final clump catalog contains the properties of regions with stellar and/or gas densities that are  at least 10 times higher than their local surroundings. 

\subsection{Examples of clumpy galaxies}

\Fig{clumps} shows a massive, $\Ms>10^{10} \ \msun$, baryon-dominated galaxy in the two datasets. 
The distribution of cold gas plus young stars (T$<1.5 \times 10^4$ K, $ \rm{age}<100 \Myr$) looks clumpy in both cases. However,  in VELA-3 there is a population of small and compact clumps that is absent in VELA-6. 
The clumps with stronger feedback look more irregular and elongated. 
The stellar distribution clearly shows dense and round stellar overdensities in VELA-3. 
The stellar disc in VELA-6, where feedback is stronger,
is significantly less clumpy and the stellar clumps are less prominent.

Young stars (and indirectly the star-forming gas) can be traced by rest-frame UV, while the near-IR traces stellar mass.
\Fig{obs} shows mock images of the same typical $z\simeq2$ galaxy shown in \Fig{clumps}. 
They are performed using Sunrise.
We created high resolution mock images for both the VELA-3 and VELA-6 simulations using the Sunrise dust radiative transfer code \citep{Jonsson06, Jonsson10}. They are a follow up to earlier generations of VELA mock images presented by \cite{Snyder15}  and \cite{Simons19}, now publicly available in the Mikulski Archive for Space Telescopes (MAST) \footnote{See https://archive.stsci.edu/prepds/vela/}. These mock images include face-on, edge-on, and random viewing orientations. 
We used the STARBURST99 stellar population models \citep{Leitherer99} with  a Milky-Way-like dust model or an SMC-like dust model. We created images using a variety of HST, JWST, and Roman broadband imaging filters, which make them suitable for comparing with observations.  We plan to submit these new mock images to MAST.
\Fig{obs}  uses the HST F606W filter for the rest-frame UV and the JWST  F277W band for the rest-frame near-IR image.
The HST image looks clumpy in both datasets. Clumps show a large variety of sizes and shapes. They all trace the same dense, star-forming regions seen in \Fig{clumps}. 
The JWST images look very different. 
In particular, bright and round clumps in VELA-3 are absent in VELA-6.
Only some clumps associated with large star-forming regions can be barely seen above the disc background. 
Therefore, the rest-frame near-IR looks less clumpy than the rest-frame UV in models with strong feedback.
This is consistent with the trends seen in other simulations \citep{Buck17} and in observations.
Images in the rest-frame optical bands look less clumpy than the UV counterparts \citep{Wuyts12, White22}.
JWST's larger diameter will allow better resolution of clumps than HST.
Our results suggest that JWST observations that provide rest-frame near-IR images of galaxies at $z\simeq2$ will be crucial for mapping the stellar distribution of clumpy galaxies and they will be able to distinguish between different feedback scenarios. 

\subsection{Clump statistics}

\begin{figure*}
	\includegraphics[width=2 \columnwidth]{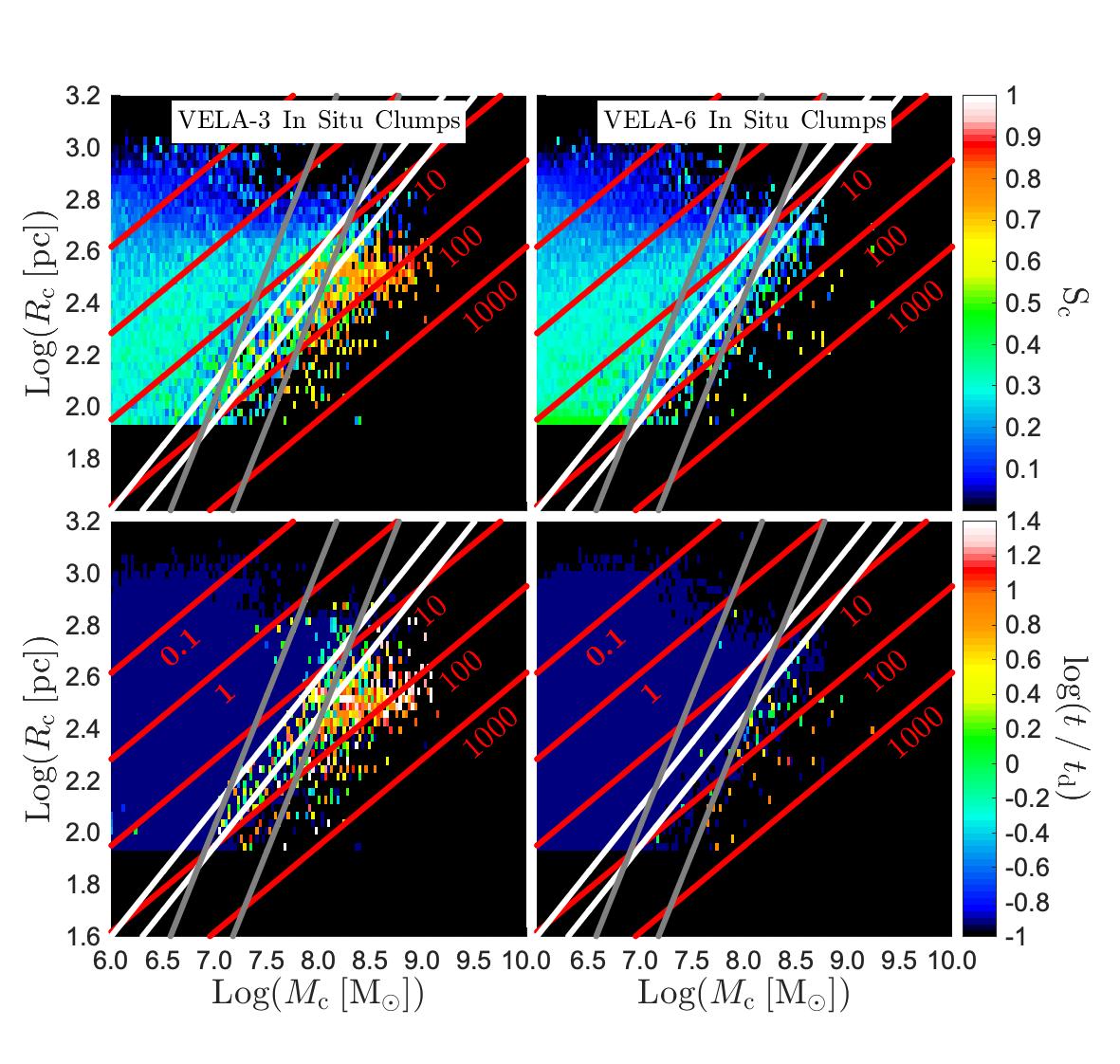}	
		\caption{Distribution of clumps in the mass-radius plane for VELA-3 (left) and VELA-6 (right). Top panels are coloured by the shape parameter, $S_c$ and bottom panels show the clump age in units of the disc dynamical time. 
		 Clumps with t = 0 were artificially set to dark blue.
		 The red lines represent constant baryonic density in units of ${\rm cm}^{-3}$.
		The grey lines represent a constant circular velocity of $V_{\rm circ}=20$ and 40 \kms, while white lines represent a constant baryonic column density of $\Sigma_{\rm bar}=200$ and $400 \msun \pc^{-2}$. VELA-6 clumps are significantly  less round. In VELA-3, migrating clumps with times longer than the disc dynamical have typically $V_{\rm circ}>20 \kms$ and $\Sigma_{\rm bar}>200 \msun \pc^{-2}$. The stronger feedback in VELA-6 pushes migrating clumps to higher values: $V_{\rm circ}>40 \kms$ and $\Sigma_{\rm bar}>400 \msun \pc^{-2}$. }
	  \label{fig:clumpsP}
\end{figure*}	

\begin{figure*}
	\includegraphics[trim={60 0 60 0},clip,width= 2 \columnwidth]{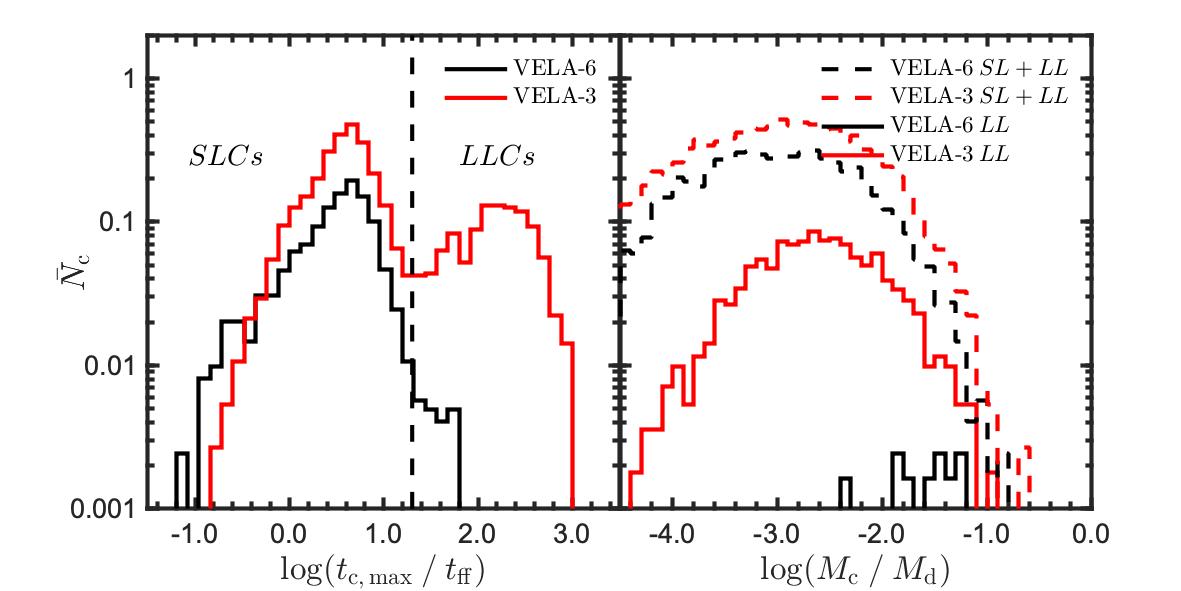}
		\caption{Distributions of clump lifetimes, $t_{\rm c,max}$, normalized to the mass-weighted clump free-fall time (left) and clump masses, normalized to the disc mass (right). Zero-lifetime clumps are excluded.
		 The number of long-lived clumps with lifetimes longer than 20 free-fall times (vertical dashed line) is drastically reduced in VELA-6. They are limited to the most massive clumps with masses similar to the Toomre mass, log($M_{\rm c}/M_{\rm d}) \simeq -1.5$. }
	  \label{fig:histo}
\end{figure*}	  

\begin{figure*}
	\includegraphics[width=2 \columnwidth]{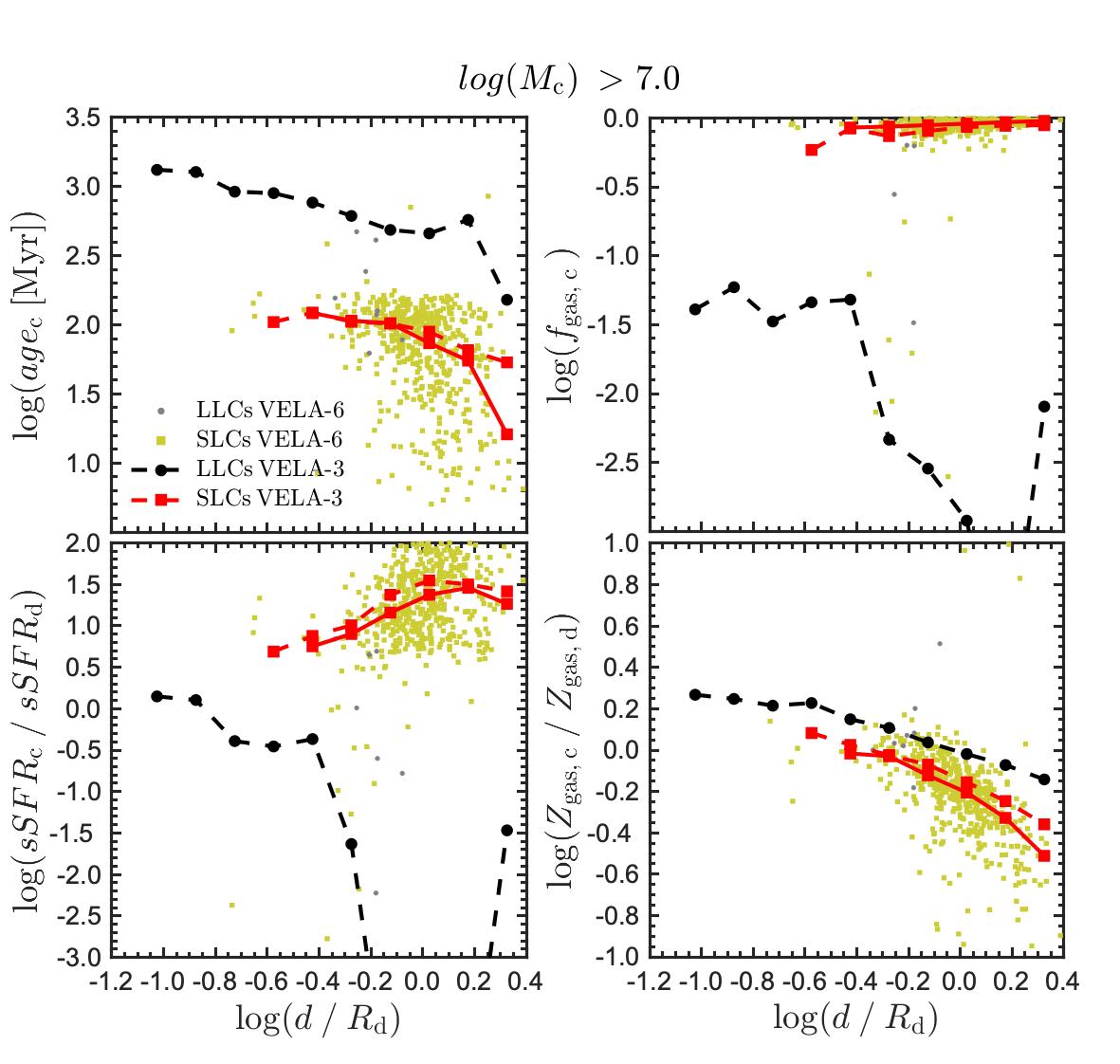}
		\caption{Gradients of clump properties with galactocentric distance for VELA-6 SLC clumps (yellow points) and LLCs (grey points). 
		Lines represent median values for VELA-6 (solid) and VELA-3 (dash).
		Top-left: mass-weighted mean stellar age; Top-right: gas fraction; bottom left clump sSFR normalized to the disc sSFR; bottom-right: clump metallicity normalized to the mean disc metallicity. In-situ clumps in VELA-6 show a steep age gradient with typical clump ages of 100-200 Myr. They are gas-rich, with higher sSFR and lower metallicities than the disc. }
	  \label{fig:gra}
\end{figure*}

We analyzed clump properties following the methods described in M17. The  clump mass, $M_c$, is the total baryonic mass (stars and gas) within the clump volume.
The clump radius, $R_c$, is defined as the radius of a sphere with the same volume as the clump. 
The shape of this volume is characterized by its inertia tensor. We define the shape parameter as $S_c=I_3/I_1$, where $I_1$ and $I_3$ are  the largest and the smallest eigenvalues of the tensor. A perfect sphere yields $Sc=1$ and a flattened oblate clump has $Sc\simeq0.5$.

Clumps are tracked through time in the simulations using their stellar particles. While tracking the clump throughout its lifetime, we use the same definition of clump time, $t$, as used in M17.
This is the time since the clump formation. When a given clump is identified for the first time, we compute $t$ as the stellar age of the clump if this is lower than the time since the previous snapshot. Otherwise, we set $t=0$. If a clump is identified at later snapshots, we increase $t$ by the corresponding time interval. $t_{\rm c, max}$ is the clump time at the last snapshot of the clump. We refer to M17 for further details. 
We only follow in-situ clumps that formed within the central galaxies. 
Ex-situ clumps are small galaxies that joined the disc through mergers. They are excluded based on their dark matter content or the birth place of their stellar particles.

\Fig{clumpsP} shows the mass-radius plane of clumps, as in Figure 6 from M17, for VELA-3 and VELA-6 in-situ clumps.
The maps are coloured by the shape parameter, $S_c$ and the clump age, $t$, normalized by the disc dynamical time, $t_d$.  
At low clump masses, the distribution of clumps is similar in both cases. This regime is dominated by clumps with $t_{\rm c, max}=0$, defined as zero-lifetime clumps (ZLC) in M17. They are mostly transient features in the density field. At higher masses, $M_c>10^7 \ \Msun$, the clumps in VELA-6 have lower baryonic densities in general, as a result of a stronger feedback.

The distribution of clump shapes is very similar, with the exception of round clumps, $S_c \ge 0.7$. 
They are not present in the runs with strong feedback, as seen in \Figs{clumps} and \ref{fig:obs}. 
The only oblate clumps with $S_c \simeq 0.4-0.6$  seen in VELA-6 correspond to the densest clumps that survive feedback. These  can retain a coherent structure in Jeans equilibrium between self-gravity, shear and feedback \citep{Dekel22a, Dekel22b}. 

\Fig{clumpsP} shows the distribution of clump age at a given snapshot with respect to the disc dynamical time.
In both datasets, there is a large fraction of transient clumps with very short ages. These clumps are quickly disrupted due to feedback right after the onset of star formation. Most of them are ZLC as described in M17. This population is also seen in other simulations with strong feedback \citep{Genel12, Oklopcic17} with lifetimes between 10-50 Myr, much shorter than the disc dynamical time.

A second population of clumps has ages longer than the disc dynamical time.  These clumps can migrate within the disc and can drive gas inflows towards the galaxy centre \citep[][M17]{DSC, CDB}. 
However, the properties of the migrating clumps differ in the two datasets.
Migrating clumps in VELA-6 are rarer and denser than in VELA-3.
They usually have $\Sigma_{\rm bar} \geq 400 \msun \pc^{-2}$ and $V_{\rm circ} \geq 40 \kms$ in VELA-6, whereas migrating clumps in VELA-3 have  $\Sigma_{\rm bar} \geq 200 \msun \pc^{-2}$ and $V_{\rm circ} \geq 20 \kms$.
It seems that migrating clumps should be denser in order to survive stronger feedback.
With respect to ages, VELA-6 migrating clumps have shorter ages, $t \le 10 \ t_d$, whereas the range of ages in VELA-3 is much broader, $t = (1 - 30 ) t_d$.
A more detailed analysis of clump evolution is discussed in a companion paper \citep{Dekel22a, Dekel22b}.

%\adrc{Add discussion: Dekel? }	

As discussed in M17, VELA-3 runs show a bimodality in clump lifetimes, $t_{\rm c,max}$ with respect to their free-fall times, $t_{\rm ff}$. We keep the same nomenclature and use the same threshold, $ 20  t_{\rm ff}$, to distinguish between short-lived clumps (SLCs) and long-lived clumps (LLCs), after excluding all ZLCs from further analysis ($t_{\rm c,max}$=0).
In VELA-6, that bimodal distribution is absent  (\Fig{histo}) and LLCs only account for an extended tail in the distribution of clump lifetimes. This is consistent with the results from isolated disc simulations with strong feedback \citep{Hopkins12b}.
Therefore, the fraction of LLC in VELA-6 is much lower than in VELA-3.
Feedback is limiting the lifetime of clumps but it also decreases the average clump densities, increasing the clump free-fall times.

The clump mass function (\Fig{histo}) is very similar among the two models
if clump masses are normalised by the disc mass, defined as all cold gas plus stars with high angular momentum within the galaxy radius (M14).
SLCs dominate by number in both cases.
Only the distribution of LLCs is radically different.
LLCs with low masses, $M_{\rm c}/M_{\rm d}  \le 0.03$, dominate the LLCs population in VELA-3.
That population disappears in VELA-6. 
The rarer LLCs in VELA-6 are limited to the most massive clumps ($M_{\rm c}/M_{\rm d} \geq 0.03$)  with masses similar to the Toomre mass. These giant clumps are massive enough to survive feedback, as steady winds generated by feedback are not expected to unbind them \citep{KrumholzDekel10}.

\begin{figure}
	\includegraphics[trim={5 5 180 5},clip,width= \columnwidth]{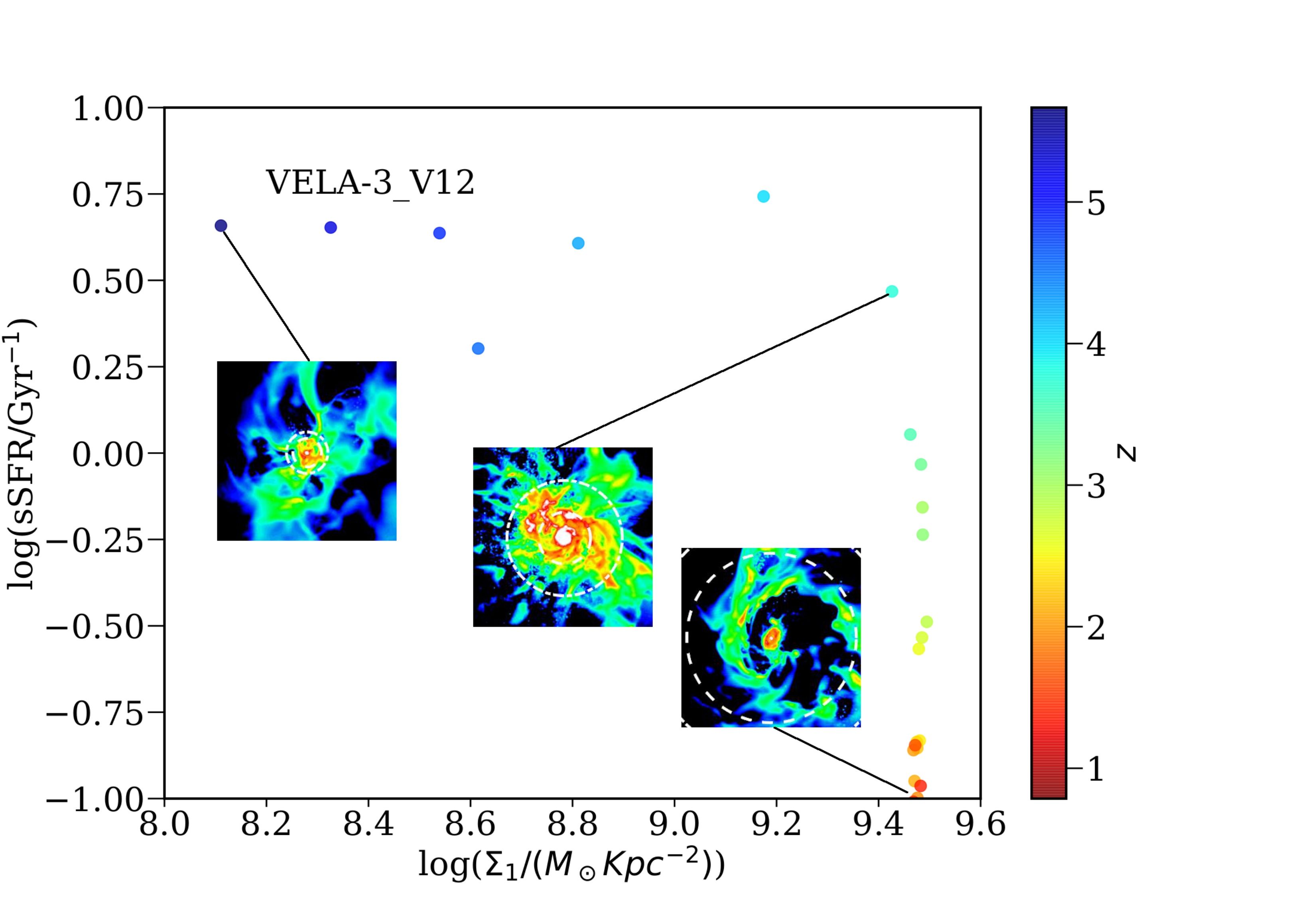}
	\includegraphics[trim={5 5 180 5},clip,width= \columnwidth]{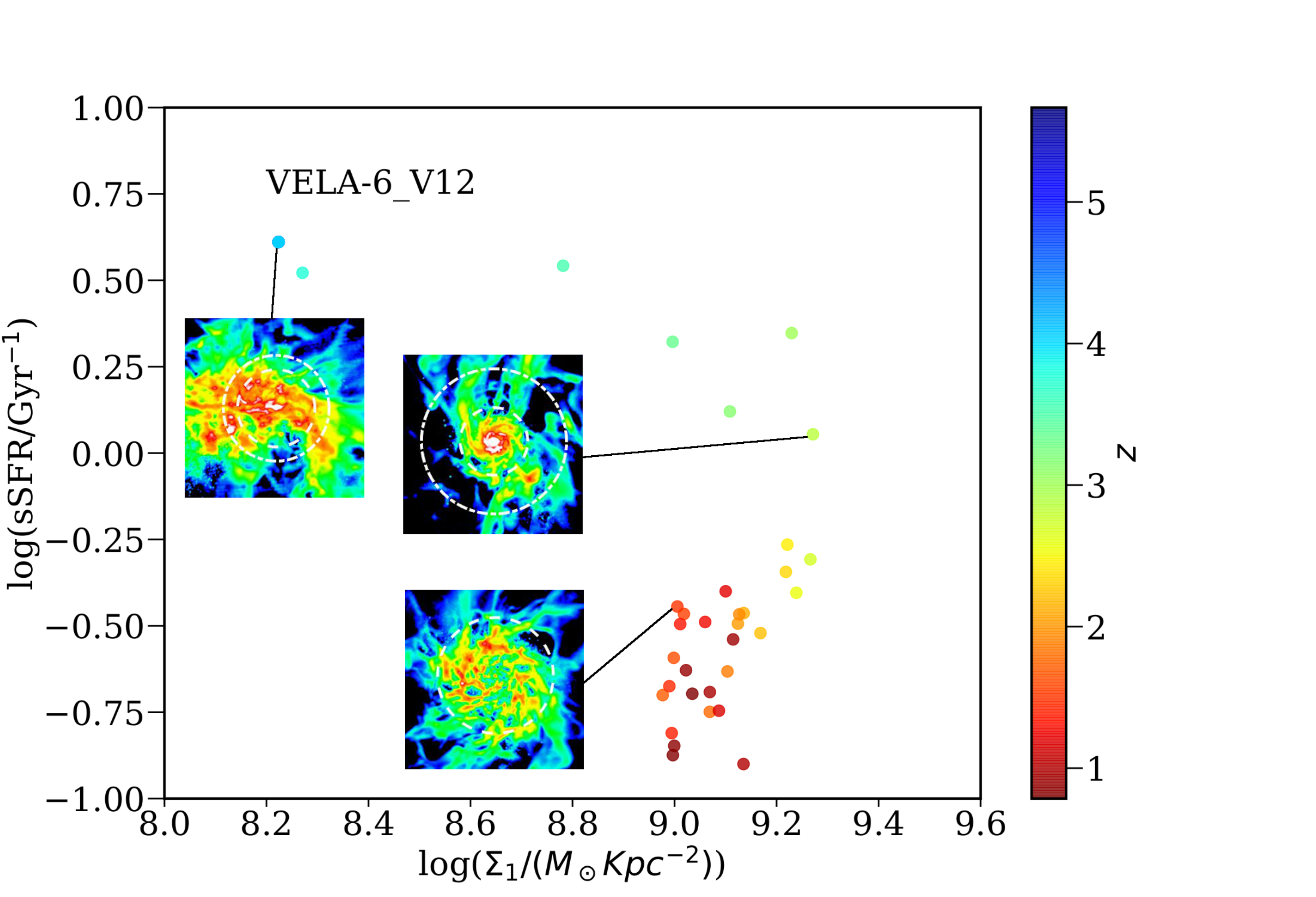}
		 \caption{Example of compaction and quenching of the galaxy V12 in the VELA-3 (top) and VELA-6 (bottom) suite in the  $\Sigma_1$-sSFR plane. The maps show the gas surface density at  the relevant phases: pre-compaction, compaction and post-compaction times with a scale of 10 kpc.}
	  \label{fig:compaction1}
\end{figure}	
\begin{figure}
	\includegraphics[trim={0 0 80 0},clip,width= \columnwidth]{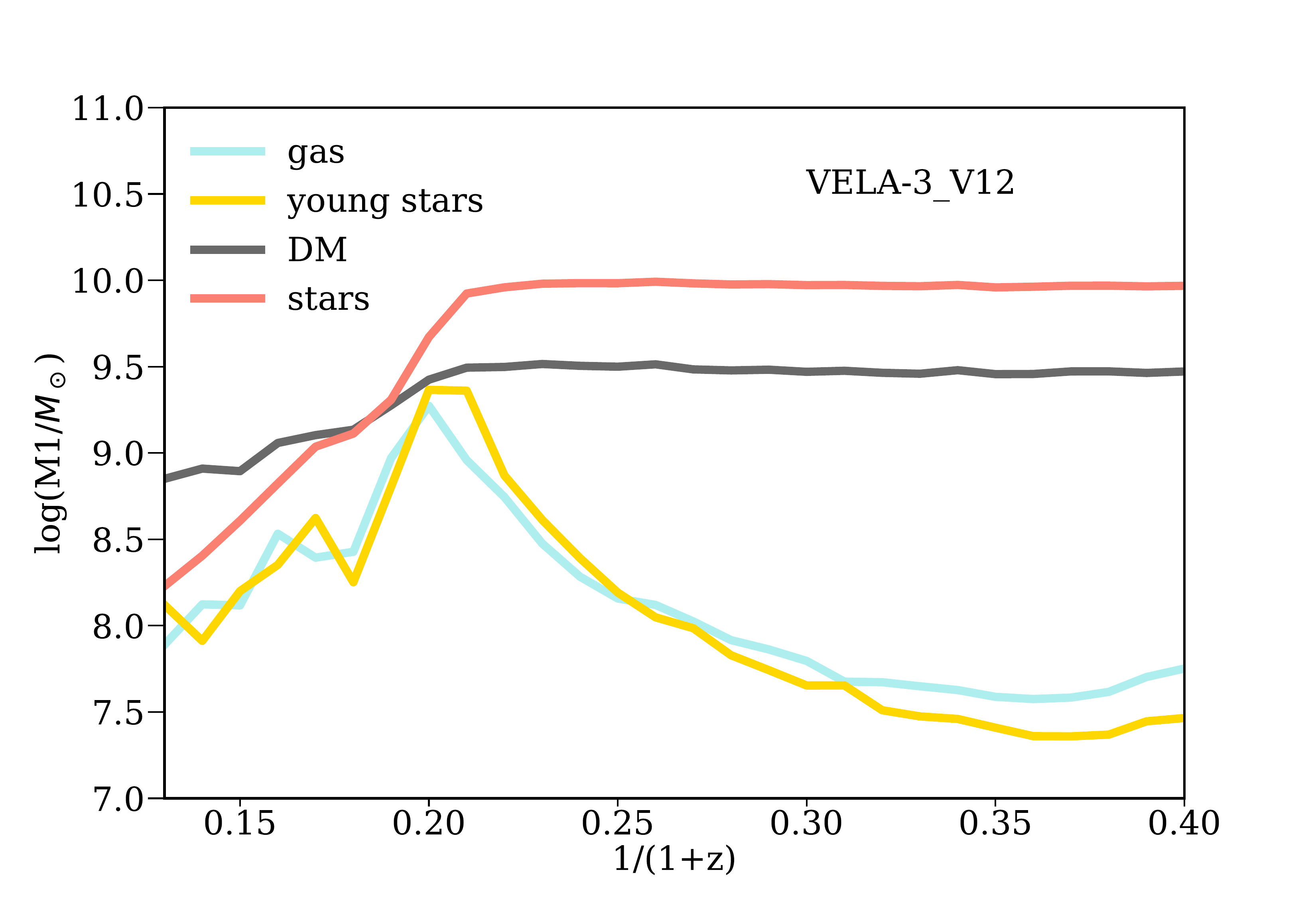}
	\includegraphics[trim={0 0 80 0},clip,width= \columnwidth]{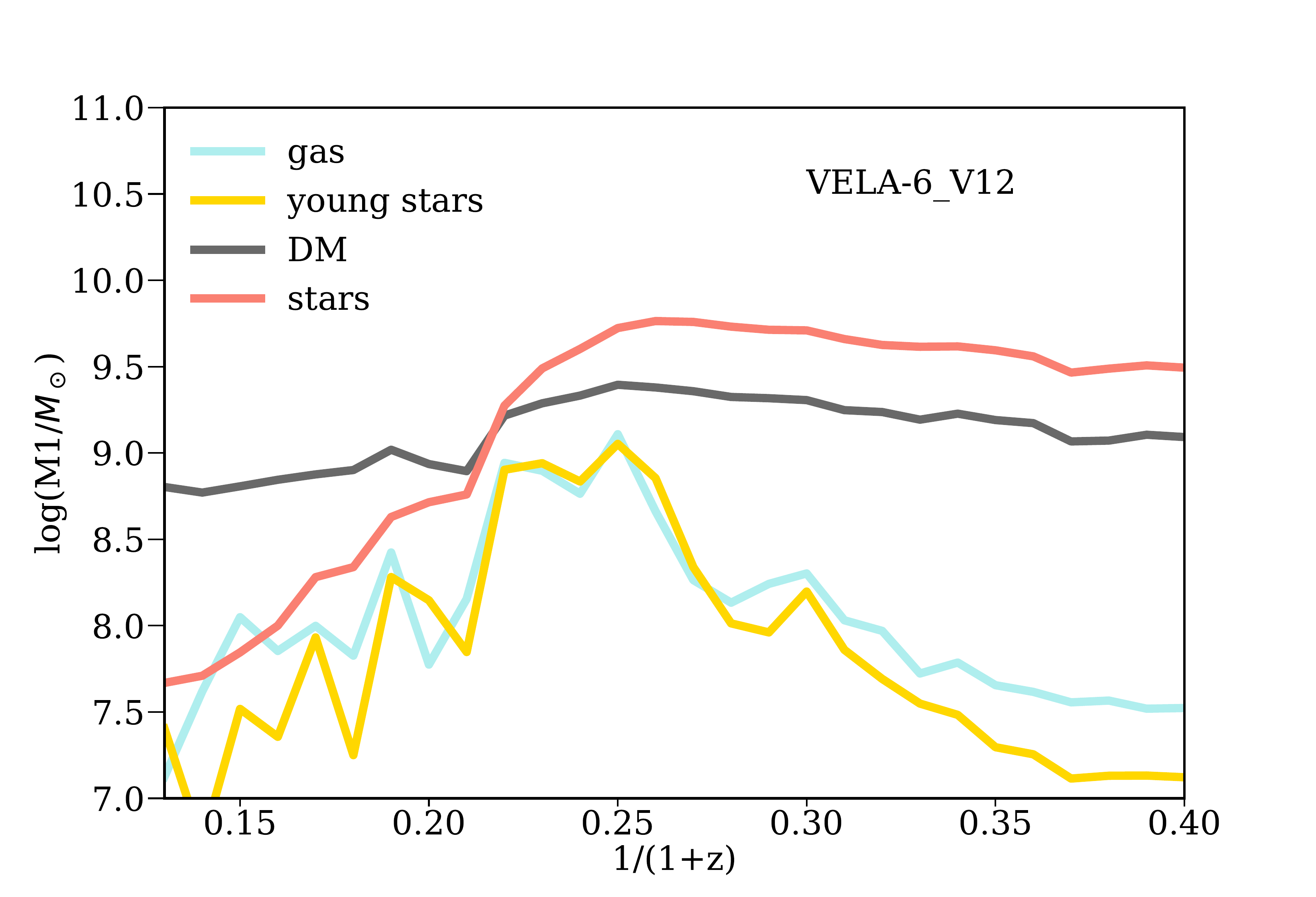}
		 \caption{Evolution of the mass within 1 kpc, M1, of the different components of the galaxy V12 in the VELA-3 (top) and VELA-6 (bottom) suite. The increase in the central gas mass starts a compaction event, weaker by a factor 2 in VELA-6.} 
	  \label{fig:compaction2}
\end{figure}

\subsection{Gradients}

\Fig{gra} compares various clump properties as a function of their cylindrical distance from the galaxy center, referred to as clump gradients, for in-situ clumps in VELA-3 and VELA-6. This is analogous to Figure 15 in M17.
The gradients of SLCs in VELA-3 and VELA-6 are relatively similar.
They show a steep age gradient with typical clump ages of 100-200 Myr. They are gas-rich, with higher sSFR and lower metallicities than the disc.
The only difference is that SLC gradients in VELA-6 show slightly lower sSFR, higher gas fraction, and lower metallicity  than the corresponding SLC gradients in VELA-3. 
The increase in feedback strength produces a decrease in the star-formation efficiency within these clumps.

LLCs show very different gradients. 
The population of old (age$_c > 300 \ {\rm Myr}$), gas poor ($f_{\rm gas,c}<0.1$) LLCs with sSFR lower than the underlying disc  and high metallicity is absent in VELA-6.
In particular, a population of  quenched clumps (log(sSFR$_c$/sSFR$_d$)<-3, 
where subscripts c and d refer to clump and disk),
which accounts for 40\% of the total population of LLCs in VELA-3 (M17) have disappeared in VELA-6. 
Feedback may prevent the formation of these stellar-rich clumps. As a results, typical  ages in VELA-6 clumps extend to  100-300 Myr, consistent with observations \citep{Wuyts12, Guo15}.
A more detailed comparison with observations requires the use of mock observations \citep{Huertas20, Ginzburg21} and 
this will be the focus of future work.

In VELA-6, there are no clumps at low galactocentric distances, $d<0.2 R_d$.
It seems that clumps in models with stronger feedback are disrupted before they reach the galaxy centre, delivering gas and stars into the inner 20\% of the disc radius.
Clump evolution and its contribution to the growth of the bulge will be discussed in follow-up papers.

%%%%%%%%%%%%%%%%%%%%%%%%%%%%%%%%%%%%%%%%%%%%%%%%%%
\section{Compaction and Quenching} 
\label{sec:C}
%%%%%%%%%%%%%%%%%%%%%%%%%%%%%%%%%%%%%%%%%%%%%%%%%%

Gas inflows within galactic discs, partially driven by migrating clumps and minor mergers, naturally trigger wet compaction and inside-out quenching \citep{DekelBurkert, Zolotov15, Tacchella16}.
\Figs{compaction1} shows an example of compaction and quenching in VELA-3 and VELA-6. 
The overall behaviour is similar in both models.
At high redshifts, the galaxy is star-forming and diffuse  (pre-compaction phase), characterized by a relatively high sSFR and low surface stellar density in the inner 1 kpc, $\Sigma_1$.
After some time, the galaxy reaches a maximum in $\Sigma_1$. 
This marks the compaction event.
The sSFR remains high during the peak of compaction.
After that maximum, the galaxy reduces its sSFR (quenching) and enters into the post-compaction phase with 
roughly constant $\Sigma_1$ within a factor of 2.
This path in the $\Sigma_1$-sSFR plane, also seen in observations \citep{Barro13}, seems robust against variations in the feedback strength.
The only noticiable difference is that compaction happens a little earlier in VELA-3 and the maximum density is higher.
Most probably this is due to a higher stellar production in VELA-3.

The evolution of the mass in the different components within the central 1 kpc is a key feature in compaction events (\Fig{compaction2}).
In the VELA-3 run, the peak in gas and SFR within 1 kpc is narrower in time than in VELA-6. The maximum gas mass is also a factor 2 lower in VELA-6.
This decreases the maximum of stellar mass by a factor of 2, because of the self-regulation by feedback.
This maximum of compaction happens after the central mass becomes dominated by stars in both cases.
It seems that feedback weakens compaction and gas inflows within the galaxy by less than a factor of 2.
After that maximum of compaction, both gas mass and SFR strongly decreases, due to gas consumption and strong outflows driven by the nuclear starburst (Z15).	
However, in VELA-6, the stellar mass and dark-matter mass decreases by a factor of 2, due to fluctuations in the gravitational potential induced by gas outflows \citep{Pontzen12}.
As a result, the stellar mass after compaction in VELA-6 is 0.5 dex lower than in VELA-3.

\begin{figure}
	\includegraphics[trim={0 0 150 0},clip,width= \columnwidth]{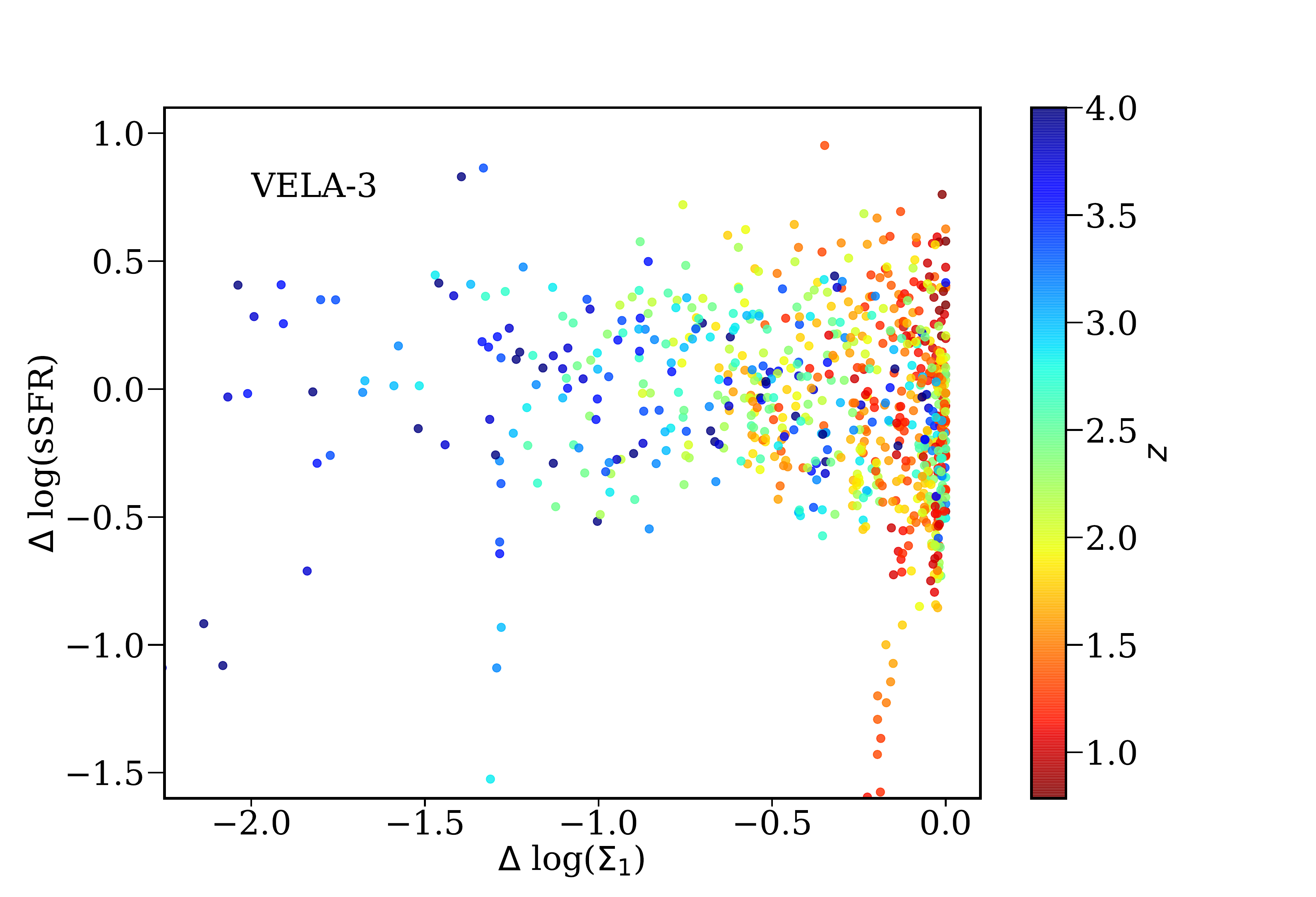}
	\includegraphics[trim={0 0 150 0},clip,width= \columnwidth]{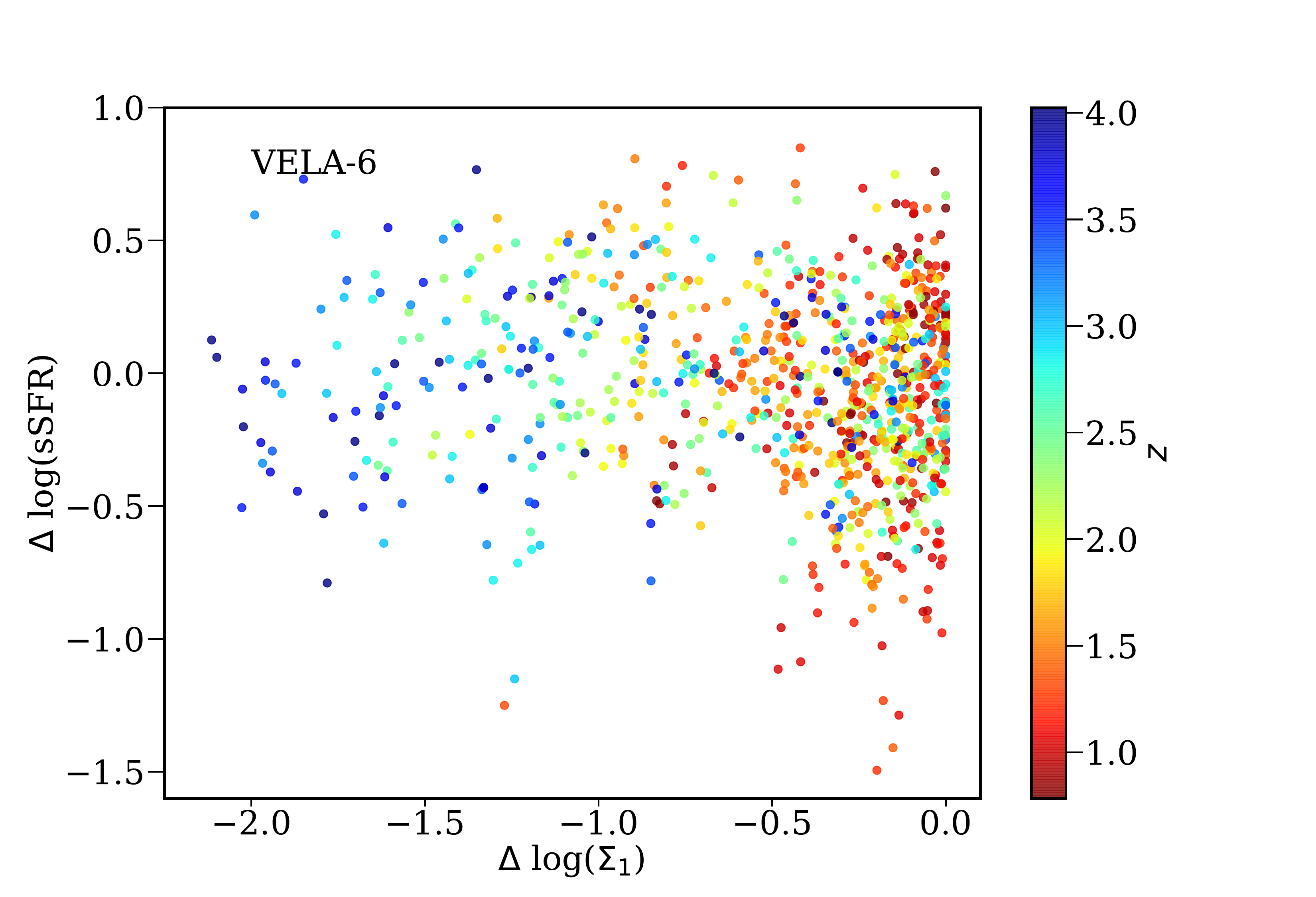}
		 \caption{log(sSFR)-log(sSFR$_{\rm MS})$ vs log($\Sigma_1$)-log($\Sigma_{1, \rm max}$) for VELA-3 (top) and VELA-6 (bottom) coloured by redshift. 
The galaxies in both models evolve along a universal L-shape track, first along a horizontal path, following the star-forming main sequence, and then, after compaction, along a vertical, quenching path.}		 
	  \label{fig:sSFR_Sigma1N}
\end{figure}

The overall evolution of high-z galaxies in the $\Sigma_1$-sSFR plane is insensitive to the strength of feedback once these properties are normalized.
\Fig{sSFR_Sigma1N}  shows this behaviour in the plane between $\Delta$log(sSFR)$=$log(sSFR)-log(sSFR$_{\rm MS}$) and $\Delta$log($\Sigma_1$)=log($\Sigma_1$)-log($\Sigma_{1, \rm max}$), where sSFR($z$)$_{\rm MS}$ marks the star-forming main sequence and $\Sigma_{1, \rm max}$ is the maximum $\Sigma_1$ value of each galaxy, as described in Z15. 
Galaxies in VELA-6 evolve around the main-sequence (horizontal branch in \Fig{sSFR_Sigma1N}), with slightly higher deviations than in VELA-3 
(Larkin et al. in prep.)
After they reach the point of maximum compaction, they evolve below the main sequence (vertical branch), initialising a process of quenching that could be completed and maintained with the aid of AGN feedback (not included in these simulations).
A recent, in-depth study of wet compaction can be found in \cite{Lapiner23}.

%%%%%%%%%%%%%%%%%%%%%%%%%%%%%%%%%%%%%%%%%%%%%%%%%%
\section{Conclusions and Discussion}
\label{sec:summary}
%%%%%%%%%%%%%%%%%%%%%%%%%%%%%%%%%%%%%%%%%%%%%%%%%%

%Summary:

We have compared two sets of cosmological, zoom-in simulations of 35 high-z galaxies with the same initial conditions but different models of stellar feedback (with and without kinetic feedback, \se{ST}).  The model with stronger feedback (VELA-6) gives the following results with respect to the previous model (VELA-3):
\begin{itemize}
\item
At a fixed halo mass and redshift, the galaxy stellar mass is reduced by a factor of $\sim$1-3, making the stellar-mass-halo-mass relation in better agreement with abundance matching results (\Fig{SMHM}).
\item
In agreement with previous works (C15), the degree of 3D elongation in a galaxy is higher for lower stellar-to-DM mass ratios within the half-light radius. Low-mass galaxies, $\Ms<10^{10} \ \msun$ at $z\ge1$, are more elongated at a fixed stellar mass in VELA-6 (\Fig{e}).
\item
The population of dense, round, compact, old (age$_c > 300 \ {\rm Myr}$), quenched (log(sSFR$_c$/sSFR$_d$)<-3), stellar (or gas poor) clumps  is absent in VELA-6.
Therefore, the distribution of stellar mass, traced by rest-frame near-IR bands, is significantly less clumpy than the distribution of star forming regions, traced by rest-frame UV (\Fig{obs}).
\item
Migrating clumps with clump times longer than the disc dynamical time have shorter stellar ages (age$_c = 100 - 200 \ {\rm Myr}$) and higher baryonic densities and circular velocities (\Fig{clumpsP}).
\item
In VELA-6, the rarer long-lived clumps with lifetimes longer than 20 free-fall times are limited to
the most massive clumps ($M_{\rm c}/M_{\rm d} \geq 0.03$)  with masses similar to the Toomre mass (\Fig{histo}).
%\item
%There are no clumps at very low galactocentric distances, $d<0.2 R_d$. 
\item
The evolution of compaction and quenching in the $\Sigma_1$-sSFR plane is similar in both models  (\Fig{sSFR_Sigma1N}).
\end{itemize}

%more discussion:
Our results suggest that JWST observations will be crucial for our understanding of galaxy evolution at high-z. 
Rest-frame near-IR images of galaxies at cosmic noon will be able to map the stellar distribution of clumpy galaxies in great detail.
%Very recent JWST images are unveiling compact stellar distributions that may come from compaction episodes \citep{Suess22}.
JWST will also observe low-mass elongated galaxies, which should dominate the number of galaxies at high-z.
The great capabilities of JWST will allow us to study these processes at even higher redshifts. 
For example, recent NIRCam observations of galaxies at $z\simeq6-8$ have shown very clumpy morphologies \citep{Chen22}.
We expect that new observational constrains on clump properties, particularly age and mass distributions, may help distinguish between different feedback models,
At the same time, the dependences of clumps on star formation models  \citep{Buck19} may add degeneracies that complicates the interpretation of these new observations. More simulations with a wide range of models are needed.

%\textcolor{magenta}{\bf [AD: Refer to the feedback puzzle, where strong feedback is needed for matching the stellar-to-halo mass ratio, but weak feedback is needed to allow massive clumps to be long-lived.  This maybe strong preventive feedback versus weak ejective/disruptive feedback.]}

The VELA project allows us to compare a large sample of zoom-in simulations performed with different models of feedback.
This unique dataset provides overall trends with feedback strength.
In \cite{Ceverino14}, \cite{Moody14} and \cite{Mandelker17},  we compared the effect of feedback in the previous suites of the VELA simulations: VELA-2 (thermal-only feedback) and VELA-3 (thermal plus radiative feedback).
The addition of radiative feedback reduces the gas mass in the high-values tail of the density distribution function. The galaxy stellar mass growth is therefore reduced by a factor $\sim$2. 
The addition of kinetic feedback reduces the stellar mass by another factor of a few.
Radiative feedback also reduces the number of long-lived clumps with respect to the thermal-only case. 
This trend continues in VELA-6. Therefore, there is a threshold for clump disruption that increases with feedback.
In VELA-2, clumps are disolved in a few free-fall times if their baryonic column density is $\Sigma_{\rm bar} < 100 \ \msun \pc^{-2}$.
The addition of radiative feedback in VELA-3 increases that threshold to  200$ \ \msun \pc^{-2}$. 
It becomes 400$ \ \msun \pc^{-2}$ with the inclusion of kinetic feedback in VELA-6.
As feedback strength increases, only denser clumps can survive for a significant amount of time.

\section*{Acknowledgements}
We thank the anonymous referee for comments that improve the quality of this paper.
We acknowledge stimulating discussions with Guillermo Tenorio-Tagle, Sergey Silich and Casiana Mu\~{n}oz-Tu\~{n}\'{o}n.
The VELA simulations were performed at the National Energy Research Scientific Computing Center (NERSC) at Lawrence Berkeley National Laboratory, and at NASA Advanced Supercomputing (NAS) at NASA Ames Research Center. DC is a Ramon-Cajal Researcher and is supported by the Ministerio de Ciencia, Innovaci\'{o}n y Universidades (MICIU/FEDER) under research grant PID2021-122603NB-C21.
NM acknowledges support from ISF grant 3061/21, from the Gordon and Betty Moore Foundation through Grant GBMF7392 and from the National Science Foundation under Grant No. NSF PHY-1748958.
AD was partly supported by ISF 861/20.

%%%%%%%%%%%%%%%%%%%%%%%%%%%%%%%%%%%%%%%%%%%%%%%%%%
\section*{Data Availability}

The data underlying this article will be shared on reasonable request to the corresponding author.

%%%%%%%%%%%%%%%%%%%%%%%%%%%%%%%%%%%%%%%%%%%%%%%%%%



\bibliographystyle{mnras}
\bibliography{Vela6_v9} % if your bibtex file is called example.bib

% Don't change these lines
\bsp	% typesetting comment
\label{lastpage}
\end{document}